\documentclass{article}

\usepackage{arxiv}

\usepackage[utf8]{inputenc} % allow utf-8 input
\usepackage[T1]{fontenc}    % use 8-bit T1 fonts

\usepackage[english,russian]{babel}
\usepackage[english,russian]{babel}
\usepackage{amsbib}
\usepackage[square,sort,comma,numbers]{natbib}
\usepackage{amsfonts}
\usepackage{amssymb}
\usepackage{hyperref}       % hyperlinks
\usepackage{url}            % simple URL typesetting
\usepackage{booktabs}       % professional-quality tables
\usepackage{amsfonts}       % blackboard math symbols
\usepackage{amsmath}
\usepackage{empheq}
\usepackage{nicefrac}       % compact symbols for 1/2, etc.
\usepackage{microtype}      % microtypography
\usepackage{cleveref}       % smart cross-referencing
\usepackage{lipsum}         % Can be removed after putting your text content
\usepackage{graphicx}
\usepackage{subfigure}
\usepackage{natbib}
\usepackage{doi}
\newtheorem{theorem}{\bf Theorem}
\newtheorem{definition}{\bf Definition}
\newtheorem{lemma}{\bf Lemma}
\newtheorem{corollary}{\bf Corollary}
\usepackage{float}

\title{Method for Evaluating the Number of Signal Sources and Application to Non-invasive Brain-computer Interface.}

\date{April 22, 2024}
\author{ \href{https://orcid.org/0000-0002-8835-6583}{\includegraphics[scale=0.06]{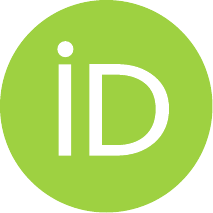}\hspace{1mm}Alexandra Bernadotte}\thanks{https://aicumene.com/.} \\
	(1) Institute of Artificial Intelligence,\\
	 M.V.Lomonosov Moscow State University;\\
	(2) International Laboratory of \\
	Algebraic Topology and Its Applications,\\
	Faculty of Computer Science,\\
	University  Higher School of Economics;\\
	(3) Rebis LLC and Aicumene LLC;\\
	(4) Neurosputnik LLC\\
	Russian Federation\\
	\texttt{a.bernadott@iai.msu.ru} \\
	\And
	\href{https://orcid.org/0000-0003-0083-6599}{\includegraphics[scale=0.06]{orcid.pdf}\hspace{1mm}Victor Buchstaber} \\
	(1) Steklov Mathematical Institute,\\
	the Russian Academy of Sciences;\\
	(2) International Laboratory of \\Algebraic Topology and Its Applications,
	\\ Faculty of Computer Science,
	\\University  Higher School of Economics;\\
	(3) Faculty of Mechanics and Mathematics,\\
	M.V.Lomonosov Moscow State University\\
	 Russian Federation\\
	\texttt{buchstab@mi-ras.ru} \\
}

\renewcommand{\shorttitle}{Method for Evaluating the Number of Signal Sources and Application to BCI.}

\hypersetup{
pdftitle={Method for Estimating the Number of Sources and Application to Non-invasive Brain-computer Interface},
pdfsubject={math, cs, physics, q-bio.NC, q-bio.QM},
pdfauthor={Alexandra Bernadotte, Victor Buchstaber},
pdfkeywords={Evaluating the number of signal sources, Brain-computer interface, BCI, Unfolding, Time series, time series unfolding method,  number of signal sources, polyharmonic signal, EEG Inverse Problem, Inverse Problem},
}

\begin{document}
\maketitle

\begin{abstract}

This paper provides a brief introduction of the mathematical theory behind the time series unfolding method.

The algorithms presented serve as a valuable mathematical and analytical tool for analyzing data collected from brain-computer interfaces.

In our study, we implement a mathematical model based on polyharmonic signals to interpret the data from brain-computer interface sensors.

The analysis of data coming to the brain-computer interface  sensors is based on a mathematical model of the signal in the form of a polyharmonic signal.

Our main focus is on addressing the problem of evaluating the number of sources, or active brain oscillators.

The efficiency of our approach is demonstrated through analysis of data recorded from a non-invasive brain-computer interface developed by the author.

\end{abstract}

\keywords{Evaluating the Number of Sources \and Brain-computer Interface \and BCI \and Unfolding \and Time Series \and Time Series Unfolding Method \and Number of Signal Sources \and Polyharmonic Signal \and Number of Sources \and EEG Inverse Problem \and Inverse Problem}

\section{Introduction}
Non-invasive brain-computer interfaces (BCIs) based on electroencephalogram (EEG) signals are currently the most prevalent due to their safety and ease of setup, in contrast to invasive BCIs and interfaces utilizing other signal modalities.

The main tasks solved with the help of non-invasive BCIs:

\begin{enumerate}
\item Real-time classification of brain states to detect mental movements lasting up to 0.5 seconds, aiding individuals with limited motor function~\cite{ref-Bernadotte, ref-Bernadotte2, ref-Bernadotte3}.

\item Classification of brain states lasting from a few seconds to hours, crucial for monitoring and detecting circulatory issues, loss of consciousness, and sleep patterns for professional (for example, to track drivers falling asleep) and medical purposes~\cite{ref-Chen}.

\item Classification of brain states over longer durations, ranging from hours to months, to identify signs of neurodegeneration, epiactivity, chronic fatigue, and other conditions for medical decision-making ~\cite{ref-Buchstaber03, ref-Luaute, ref-Bernadotte4, ref-Saltykova}.
\end{enumerate}

To effectively address real-world problems using BCIs and gain insight into brain function, overcoming the challenge of the inverse problem is essential, involving the reconstruction of signal sources and their connections from BCI data~\cite{ref-source1, ref-source2, ref-source3, ref-source4}.

Solving the inverse problem of reconstructing signal sources entails estimating the number of such sources. This paper proposes a solution for estimating the number of signal sources in the brain based on BCI data, utilizing the time series unfolding method introduced by V. M. Buchstaber in 1994~\cite{ref-Buchstaber, ref-Buchstaber2}.

\section{Brain Signal Model}\label{BrainSignalModel}

When developing techniques for analyzing brain signals acquired through non-invasive Brain-Computer Interfaces (BCIs), it is hypothesized that the brain comprises signal generators such as harmonic oscillators.

From a morpho-functional perspective, brain oscillators are clusters of interconnected neurons. Each neuron acts as an individual oscillator with its unique carrier frequency. These neurons can shift their oscillation phase in response to external stimuli from other neurons or sensory receptors, such as cone cells in the eye or taste buds.

To simplify the analysis, clusters of neurons functioning synchronously or almost synchronously are considered as a single oscillator. However, this simplification may limit the method's capabilities and introduce bias.

The structural features of an oscillator consisting of a cluster of neurons lead to the fact that we have to work neither with periodic nor quasi-periodic functions approximating the oscillator signal. We have to analyze the almost periodic functions: \\

\begin{definition} \label{def:1}A function $f$ is almost periodic if any sequence of functions $f(t + n \Delta t)$, $n \to \infty$, converges uniformly in $t$ to some periodic function.
\end{definition}

\begin{figure}
	\centering
	\includegraphics[width=200pt]{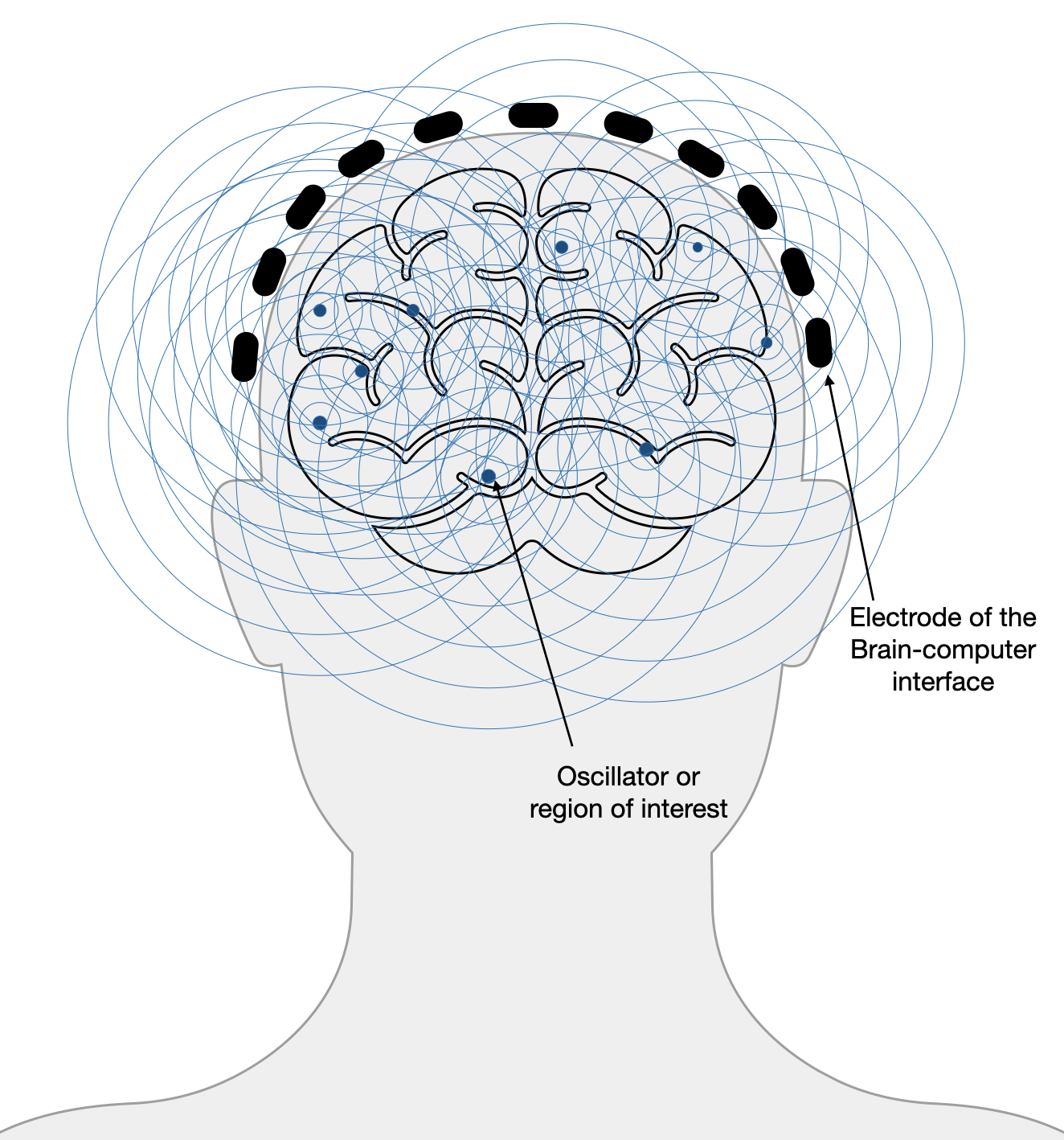}
	\caption{Schematic representation of brain oscillators and BCI sensors.}
	\label{figEEG01}
\end{figure}

The brain's electromagnetic signals, representing the sum of signals from various brain oscillators or regions of interest, are transmitted to BCI sensors, typically located on the scalp. See Fig. ~\ref{figEEG01}.

The model used to describe the signal reaching the BCI sensors is based on 
a second-order differential equation (the heat conduction equation, diffusion, and wave equation). Among the solutions of this equation, we are interested in those that are represented by a finite sum:

\begin{equation} \label{eq:1}
h(t) = \sum^{m}_{k=1} b_k sin(\omega_{k}t + \varphi_{k}),
\end{equation}
$b_k, \omega_{k}, \varphi_{k}$ -- constants.

Within this model, the number of oscillation sources is equal to $m$.

The model described as (\ref{eq:1}) is characterized by a functional equation of the form:

\begin{equation} \label{eq:1_1}
h(t + \tau) = \sum_{k=1}^{2m} h_k^1(t)h_k^2(\tau),
\end{equation}

The function~(\ref{eq:1}) is used in the analysis of the signals received by multiple sensors to solve the inverse problem of estimating the number of sources (oscillators), including their characteristics, and relationships.

The problem of representing a real signal by functions of the form~(\ref{eq:1}) is widely known. Various methods exist to estimate the parameters  $b_k, \omega_{k}, \varphi_{k}$ in this model.

 but the accuracy of these estimates is heavily influenced by the number $m$ estimation.

The accuracy of these estimates is heavily influenced by the estimate of the number $m$, especially in the case of a noisy signal.\\

The method of multidimensional unfolding of time series is described as an approach to estimating the number of sources and validating this estimation before employing more complex algorithms.

\section{Time series multidimensional unfolding method}

The discussed time series multidimensional unfolding method will be abbreviated as the TSMU method. 

The TSMU method is based on the transition from a time series to its unfolding -- a piecewise linear curve in $\mathbb{R}^n$ \cite{ref-Buchstaber}.

%The MPRS method for signal analysis allows using: firstly, electronic templates, and secondly, databases of reference signals.

%As a reference, we can use both a model signal and data marked by experts (for example, doctors) to identify typical features of the signal relevant to the task.

The TSMU method allows using a database of model signals (reference time series or templates). The database of model signals is formed both on the basis of theoretical research and on experimental studies. For example, such a database is a set of characteristic signals that have been certified within the framework of evidence-based medicine for the purpose of identifying typical signs of medical condition.

\subsection{Basic concepts and definitions of TSMU method}

\begin{definition} \label{def:2}A time series $f = (f_1, ..., f_N)$ is a sequence of values $f_k$ of a real-valued function $f(t)$, where $f_k = f( t_k), t_k = t_1 + k\Delta t, k \in \{0, ..., N\}$ and $\Delta t$ is a fixed parameter.\\
\end{definition}

BCI's data is obtained in the form of time series.\\
The time series $f$ is associated with a piecewise linear curve $X_f$.

\begin{definition} \label{def:3}
An $n$-dimensional unfolding $X_f = X_f(N, n)$ of a time series in $R^n$ is an oriented piecewise linear curve $X_f$ obtained by successively connecting vectors $X_1, X_2, ..., X_p$ (nodes of the curve $X_f$), where $X^{T}_i = (f_i, f_{i+1}, ..., f_{i+n-1})$, $p = N - (n -1)$.\\
\end{definition}

The set $T^N$ of all time series $f_1, ..., f_N$ is identified with the Euclidean space $\mathbb{R}^N$: $T^N \approx \mathbb{R}^N $.

The space $M(n, p)$ of all piecewise linear curves with $p$ nodes in $R^n$ is identified with $\mathbb{R}^{np}$: $M(n, p) \approx R^{np}$.

A piecewise linear curve $X_f$ of a time series unfolding can be presented as a matrix, where the columns of the matrix are $n$-dimensional vectors or nodes of the curve $X_f$: $X_1, X_2, ..., X_p$. The resulting matrix of $p$ columns and $n$ rows is a Hankel matrix -- a matrix of the form~(\ref{eq:2}).

\begin{equation} \label{eq:2}
H_f =
\begin{bmatrix}
f_{1} & f_{2} & f_{3} &\cdots & f_{N - n + 1}\\
f_{2} & f_{3} & f_{4} &\cdots & f_{N - n + 2}\\
f_{3} & f_{4} & f_{5} &\cdots & f_{N - n + 3}\\
\vdots  & \vdots  & \vdots  & \ddots & \vdots  \\
f_n & f_{n+1} & f_{n+2} & \cdots & f_{N}
\end{bmatrix}
\end{equation}
The description of a time series in the form of a Hankel matrix allows one to use the theory of these matrices in the analysis of time series.

\subsubsection{Ranks of Time Series' Unfoldings}

\begin{definition} \label{def:4}
A piecewise linear curve $X \in M(n, p)$, where $M(n, p)$ is the space of all piecewise linear curves with $p$ nodes in $\mathbb{R}^n$, has rank $r$ if it lies in an $r$-dimensional affine subspace $L \subset \mathbb{R}^n$.
\end{definition}

\begin{definition} \label{def:5}
A piecewise linear curve $X \in M(n, p)$ with nodes $\{X_1, ..., X_p\}$ belongs to the $\varepsilon$-neighborhood of the $r$-dimensional affine subspace $L \subset \mathbb{R}^n$ if $\Sigma^p_{i=1} ||L - X_i||^2 < \varepsilon$, where $||L - X_i||$ is the distance of the vector $X_i$ from the subspace $L$.\end{definition}

\begin{definition} \label{def:6}
A time series $f$ has $\varepsilon$-rank $r$ if its piecewise linear curve $X_f \in M(n, p)$ lies in the $\varepsilon$-neighborhood of the $r$-dimensional subspace $L$.\end{definition}

\begin{definition} \label{def:7}
The scattering matrix $W = (w_{ij})$ of a set of vectors $X$, $X= \{X_1,\ldots, X_p\}$, where $X_k \in \mathbb{R}^n$, is the Gram matrix of the set $\widetilde{X} = \{\widetilde{X_k}\}$, where $w_{ij} = <\widetilde{X_i}, \widetilde{X_j}>$ is the standard scalar product,\\ $\bar{X} = \frac{1}{p} (X_1 + \ldots+ X_p)$, $\widetilde{X_k}= X_k - \bar{X}$.\end{definition}

The matrix $W$ is symmetric and non-negative definite.\\

Let $\lambda_1 \geq \lambda_2 \geq ... \geq \lambda_n \geq 0 $ be the set of eigenvalues of the matrix $W$ and $v_1,..., v_n$ be the orthonormal set of eigenvectors of this matrix, where $v_k$ is the eigenvector corresponding to $\lambda_k$, which is called the $k$-th principal component of the set $X$.\\

\begin{definition} \label{def:8} The scattering matrix of a curve $X \in M(n, p)$ is the scattering matrix $W$ of the set of its nodes $X = \{X_1, ..., X_p\}$.
\end{definition}

It follows from the principal component theory:

\begin{lemma}\label{thm1}
The rank of the curve $X \in M(n, p)$ does not exceed $r$ $\iff$ $\lambda_i = 0$ for all $i > r$, where $\lambda_1 \geq \lambda_2 \geq ... \geq\lambda_n \geq 0 $~--~eigenvalues of the scattering matrix $W$. In this case, the $\varepsilon$-rank of the curve $X$ does not exceed $r$ $\iff$
\begin{equation} \label{eq:3}
\sum_{i=r+1}^n \lambda_i < \varepsilon.
\end{equation}\\
\end{lemma}

The curve $X \in M(n, p)$ has rank $r$ $\iff$ $X = \bar{X} + V(r)\Big(V(r)^T X\Big)$, where $V(r)$~--~$(r\times n)$ is a matrix whose columns $v_1, ... , v_r$ are the eigenvectors of the scattering matrix of the curve $X$.

\subsubsection{Time Series Reconstruction} 
 
There is a filtration in the space $M(n, p)$: $M^1(n,p) \subset M^2(n,p) \subset ... \subset M(n, p)$, where $M^r(n, p)$ is the subspace of piecewise linear curves of rank $r$.

\begin{definition} \label{def:10}
A curve $Y_{*}(r) \in M^r(n, p)$ is a projection of a curve $Y \in M(n, p)$ if:
% if the following holds:
\begin{equation} \label{eq:7}
||Y - Y_{*}(r)||^2 = \min_{\hat Y \in M^r(n,p)} ||Y- \hat Y ||^2.
\end{equation}
\end{definition}

In the space $M(n, p)$, we defined the subspace of the time series $T^N$ as the subspace of the unfoldings of these series.

\begin{definition}\label{def:11}
A time series $\hat f = (\hat f_1, ..., \hat f_N)$ is called the projection of the curve $ Y \in M^r(n, p)$ onto the space of time series $T^N$ if:
\begin{equation} \label{eq:8}
||Y -X_{\hat f}||^2 = \min_{f \in T^N} ||Y - X_f||^2,
\end{equation}
where $X_{\hat f}, X_f$ are the unfoldings of time series $\hat f$ and $f$.
\end{definition}

\begin{theorem}\label{thm2}
For a curve $Y \in M(n,p)$, its projection $\hat f(Y)$ onto the time series space $T^N$ has the form $\hat f(Y) = (\hat f_1, ..., \hat f_N)$, where :

\begin{equation} \label{eq:8}
\hat f_k =
\begin{cases}
\frac{1}{k} \sum_{i=1}^{k}\alpha_{i, k-i+1}, & 1 \leq k \leq n, \\
\\
\frac{1}{n} \sum_{i=1}^{n}\alpha_{i, k-i+1}, & n \leq k \leq p, \\
\\
\frac{1}{N- n +1} \sum_{i=1}^{N-k+1}\alpha_{i+kp, p-i+1}, & p \leq k \leq N, \\
\end{cases}
\end{equation} where $a_i \neq 0$.

\end{theorem}

\subsection{Evaluating the Number of Sources (Oscillators)}

\begin{corollary}[Characteristic property]
If the model function $f(t)$ satisfies the addition theorem (see formula \ref{eq:1_1}), then the unfolding $X_f$ of the time series obtained from it with a step $\Delta t$ lies in $\mathbb{R}^{2m}$, where $2m < n$.
\end{corollary}

That is, if there is a hypothesis that the signal is polychromatic and consists of $k$ oscillators, then its unfolding lies in the space of $2k$ dimension.

Consequently, the value of the unfolding's rank gives an estimate of the number of components of the time series in the model~(\ref{eq:1}). The projection of the unfolding onto the plane of pairs of principal components gives information about these components.

\subsubsection{Algorithm for Evaluating the Number of Sources (Oscillators)}

The algorithm consists of the following main blocks:

\begin{enumerate}
\item Obtaining a discrete signal from the BCI in the form of real-valued time series $f^1  = (f^1_1, ..., f^1_N), ..., f^s  = (f^s_1, ..., f^s_N)$  of a fixed length $N$, where $s$ is the number of BCI sensors. The sampling frequency of the BCI's analog-to-digital converter, according to Kotelnikov's theorem (1933) (Nyquist–Shannon sampling theorem, 1949), imposes a limitation on the signal frequencies available for analysis~\cite{ref-Kotelnikov, ref-Shannon}.

\item Obtaining a time series unfolding in the form of a curve $X_f$ with nodes in the form of the Hankel matrix. If there is no a priori information on the expected number of signal sources, the unfolding window $n$ should be chosen as maximum possible. The maximum initial $n = [\frac{N + 1}{2}]$. This yields a Hankel matrix of size $n \times p, p = N - n + 1$, with rank $r \leq n$.

\item Forming the scattering matrix $W$ of $X_f$ and calculating the eigenvalues of the scattering matrix $W$.

\item Evaluating the rank of the curve $X_f$ as the maximum number of eigenvalues while the residuals can be discarded.

%\item Evaluating the rank of $X_f$. If there is a priori information about the eigenvalues of the scattering matrix $W$ that can be neglected, then the $\varepsilon$-rank of $X_f$ is estimated.

%At this step, an expert assessment may be required, which will take into account the peculiarity of the problem. To do this, the time series is reconstructed and the reconstructed signal is compared with the original one as follows:
This step is taking into account a priori information and expert assessments:

\begin{enumerate}
\item Filtering the scan data $X_f$ using the subspace $M^r(n, p)$ spanned by $r$ eigenvectors, $r  \in \{1, ..., n\}$. The output of this step is $r \times p$-matrices that contain the nodes of the projections of the unfoldings $X_f$ onto the subspaces spanned by $r$ eigenvectors.

\item Reconstruction of the time series from $r \times p$-matrices using formula~(\ref{eq:8}). The output of this step is a time series, which unfolding lies in the $r$-dimensional subspace $M^r(n, p)$. The resulting $r$-th time series is considered the $r$-th model (approximation) of the original series $f$.

\item  Comparison of the reconstructed $r$-th time series with the original time series. The difference between the series is assessed according to the noise criteria. This step solves the minimax problem, which chooses $r^{*}$ based on a compromise: we get the minimum $r$ while the residuals satisfy the noise criteria.
%If the difference between the original and reconstructed series is noise, then $r$ (or $\varepsilon$-rank) is fixed as the desired rank.
\end{enumerate}

The rank $r$ (or $\varepsilon$-rank) corresponds to the number of signal sources: $m = \frac{r}{2}$, which form the signal recorded at a certain sensor.
The output of the algorithm is the number of active oscillators $m$ for a given sensor.
\end{enumerate}

\section{Results of applying the TSMU algorithm to the model and BCI data}
The analysis used model time series, the choice of which was justified by the presented theory, and data obtained from the author's BCI.\\

%\section{Practical part}
%We will demonstrate the proposed algorithm on the BCI data, as well as on the model data.
\subsection{Comparative analysis of the TSMU algorithm on model and BCI  data}

When analyzing, the choice of the projection of the unfolding onto the planes spanned by the pairs of eigenvectors of the scattering matrix affects the decision-making when evaluating the number of independent harmonic components of the signal.

%When analyzing, the choice of projection onto a certain pair of eigenvectors affects the decision-making on the presence of a number of independent harmonic components of the signal.

The interpretation of these projections is based on the theory of the TSMU model signals.

\begin{figure}[H]
   \centering
       \subfigure[The model function $g(t) = sin(\frac{\pi}{8}t)$.]{\includegraphics[width=0.49\linewidth]{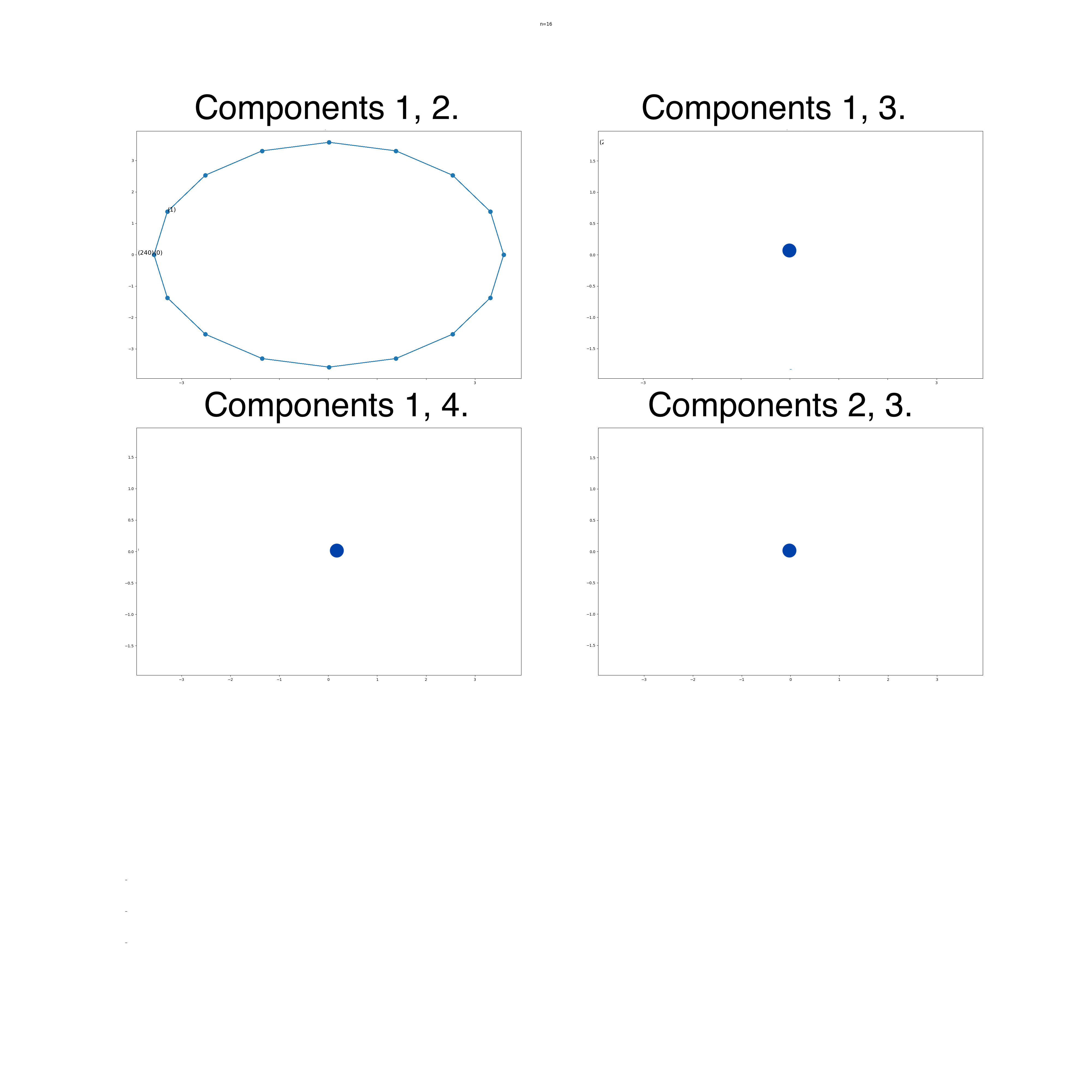}}
   \subfigure[The model function $g(t) = sin(\frac{\pi}{8}t) + \frac{1}{2} sin(\frac{2\pi}{8}t)$.]{\includegraphics[width=0.49\linewidth]{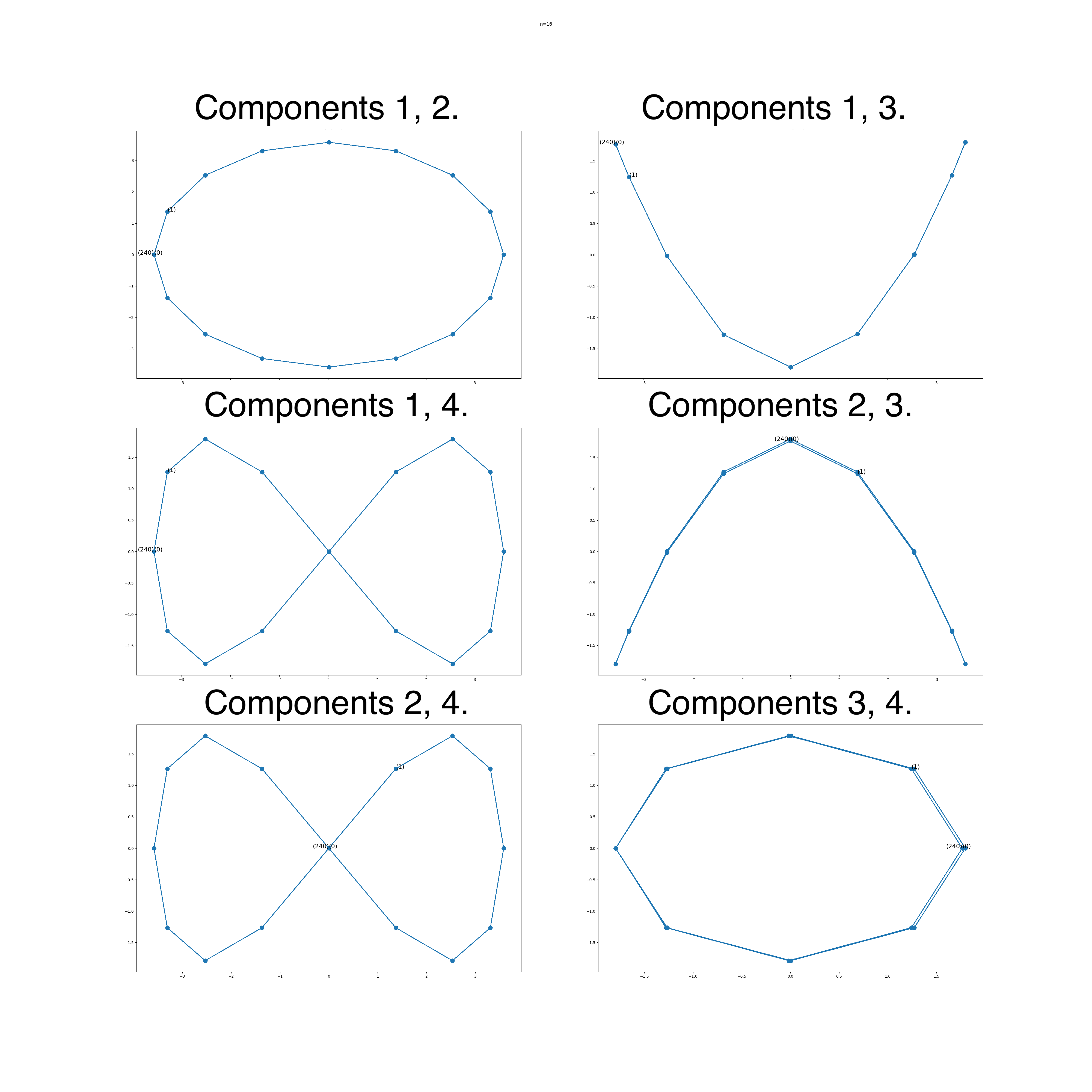}}
   \caption{The model functions projections onto the planes of the principal components. $ T = 16, T = 16, \Delta t = 1, t_1 = 0, N = 256$.}
   \label{fig_ber_1}
\end{figure}

According to the TSMU theory, the projection of the unfolding of a monochromatic signal onto the plane of the first two principal components is a circle. 
While it's projection on the planes spanned onto pairs of eigenvectors with numbers greater than 2 is a point. See Fig.~\ref{fig_ber_1}(a).

%We will consider the use of model functions as reference signals for analyzing BCI data.

%According to the unfolding theory, the projection of the unfolding of a monochromatic signal onto a plane yields a circle, if the plane is chosen correctly.

We show the application of this fact in the model function $g(t) = \sin(\frac{2\pi}{16}t) + \frac{1}{3}\sin(\frac{2\pi}{8}t)$. In Fig.~\ref{fig_ber_1}(b), pay attention to the projections onto components 1-2 and 3-4 of the model function.

\begin{figure}[H]
   \centering
       \subfigure[Model function $g(t)$.]{\includegraphics[width=0.49\linewidth]{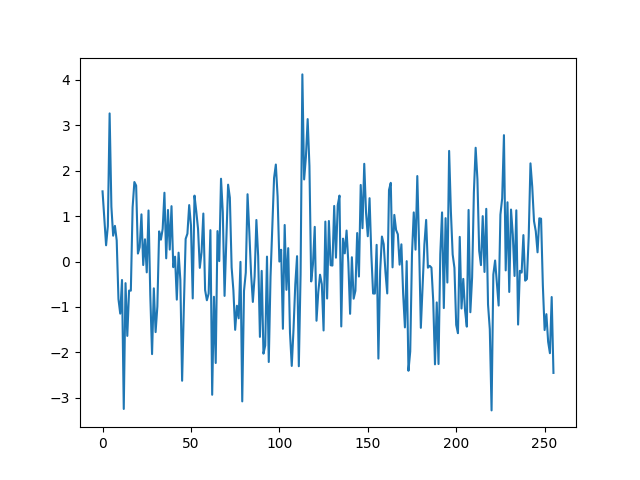}}
   \subfigure[BCI data.]{\includegraphics[width=0.49\linewidth]{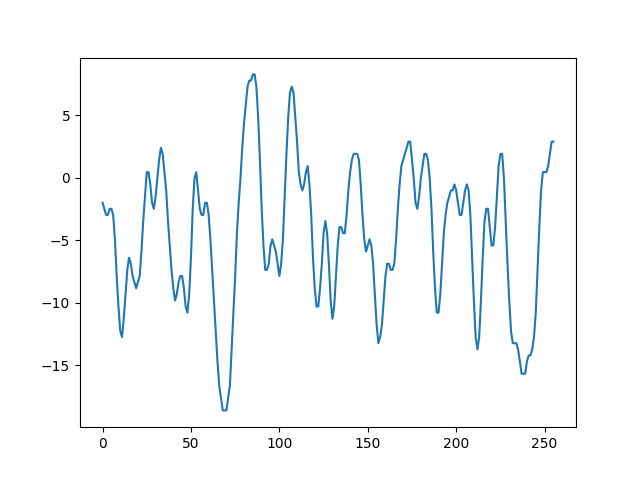}}
   \caption{ The model function
$g(t) = \sin(\frac{2\pi}{16}t) + \frac{1}{2}\sin(\frac{2\pi}{8}t) +  \frac{1}{3}\sin(\frac{2\pi}{6}t) + \frac{1}{4}\sin(\frac{2\pi}{5}t) + \varepsilon (t)$, where $\varepsilon (t)$ is set by the Gaussian noise generator ($D = 1, E = 0$) and the BCI signal. $n = 16, \Delta t = 1, t_1 = 0, N = 256.$}
   \label{fig_ber_2}
\end{figure}
 
 \begin{figure}[H]
   \centering
           \subfigure[Unfolding of the model function $g(t)$.]{\includegraphics[width=0.49\linewidth]{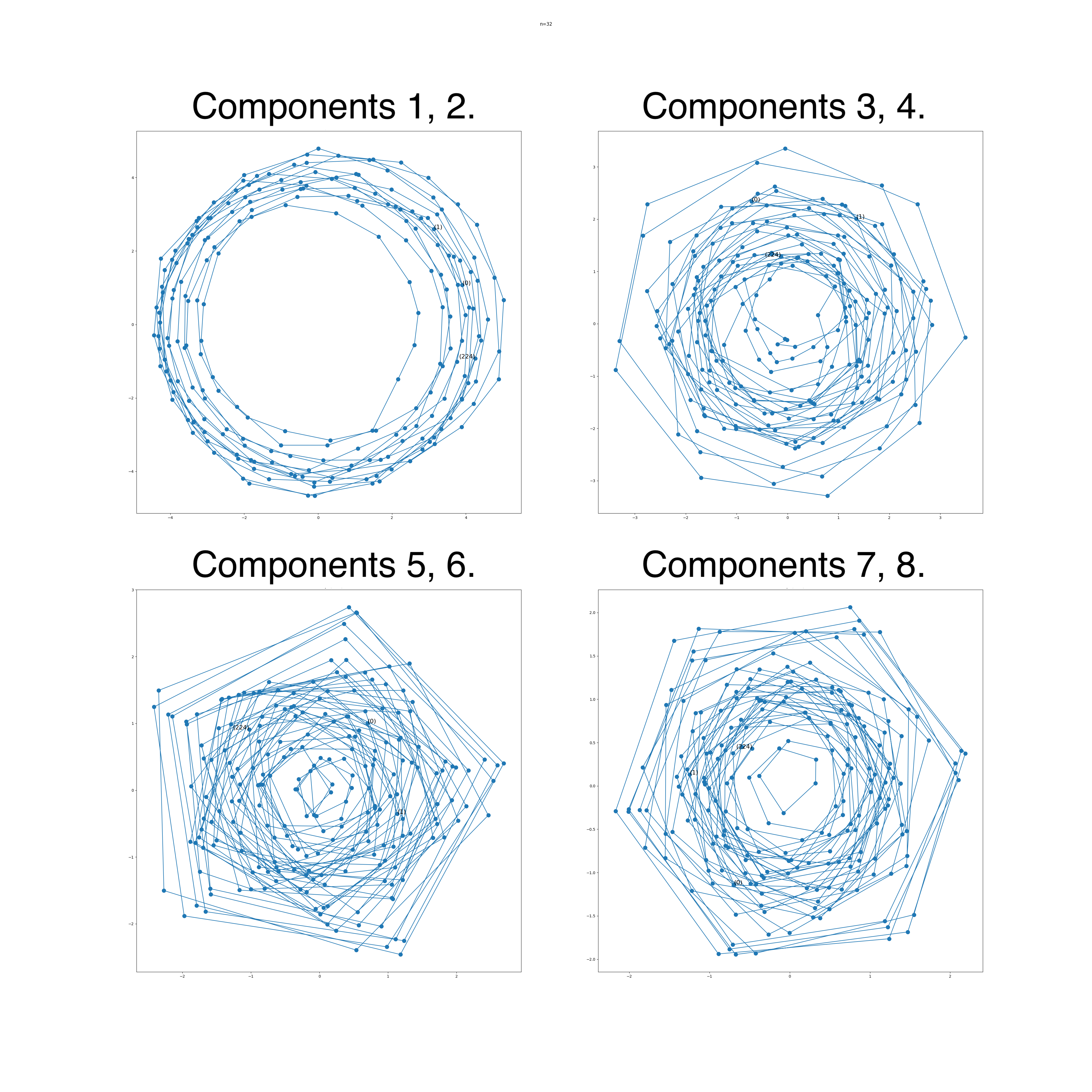}}  
           %{N256_sin_2pi_16_1_3_sin_2pi2_16_N_pca.png}}
               \subfigure[Unfolding of the BCI data.]{\includegraphics[width=0.49\linewidth]{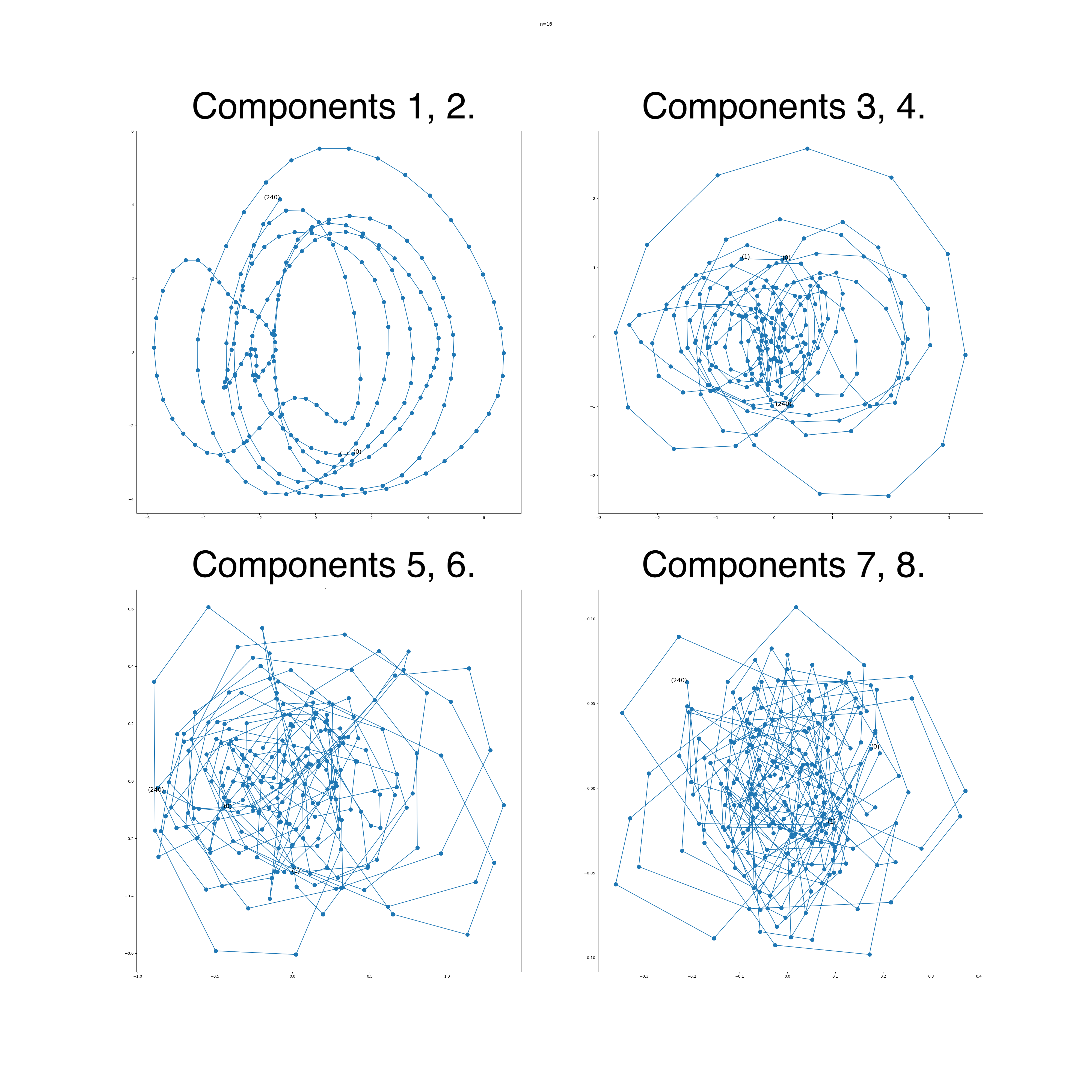}}
   \caption{Unfoldings of the model function
$g(t) = \sin(\frac{2\pi}{16}t) + \frac{1}{2}\sin(\frac{2\pi}{8}t) +  \frac{1}{3}\sin(\frac{2\pi}{6}t) + \frac{1}{4}\sin(\frac{2\pi}{5}t) + \varepsilon (t)$, where $\varepsilon (t)$ is set by the Gaussian noise generator ($D = 1, E = 0$) and the BCI signal. $n = 16, \Delta t = 1, t_1 = 0, N = 256.$ Projections of time series unfoldings onto 1-2, 3-4, 5-6 and 7-8 eigenvectors' plans. }
   \label{fig_ber_3}
\end{figure}

To demonstrate the method of Determining the Number of Sources (Oscillators) in the BCI signal, we further use, as a reference signal, a model function of the form: $g(t) = \sin(\frac{2\pi}{16}t) + \frac{1}{2}\sin(\frac{2\pi}{8}t) + \frac{1}{3}\sin(\frac{2\pi}{6}t) + \frac{1}{4}\sin(\frac{2\pi}{5}t) + \varepsilon (t)$, where $\varepsilon (t)$ is given by the Gaussian noise generator ($D = 1, E = 0$). See Fig.~\ref{fig_ber_2}(a). The signal from the BCI is a recording from the left frontal lobe in a calm state of a healthy user, 1 second long and with a sampling rate of 256 Hz. See Fig.~\ref{fig_ber_2}(b).\\

An analysis of projections of the model signal and BCI signal unfoldings onto planes formed by 1-2, 3-4, 5-6, 7-8 pairs of eigenvectors (ordered by decreasing order of the corresponding eigenvalues by vectors) shows characteristic common properties. See Fig.~\ref{fig_ber_3}.

%\begin{figure}[thpb]
 %  \centering
  %         \subfigure[Reconstruction of $g(t)$.]{\includegraphics[width=0.9\linewidth]{pictures/Figure_4_model.png}}  
           %{N256_sin_2pi_16_1_3_sin_2pi2_16_N_pca.png}}
   %            \subfigure[Reconstruction of the BCI data.]{\includegraphics[width=0.9\linewidth]{pictures/c_18.png}}
   %\caption{Time Series Reconstruction of
%$g(t) = \sin(\frac{2\pi}{16}t) + \frac{1}{2}\sin(\frac{2\pi}{8}t) +  \frac{1}{3}\sin(\frac{2\pi}{6}t) + \frac{1}{4}\sin(\frac{2\pi}{5}t) + \varepsilon (t)$, where $\varepsilon (t)$ is set by the Gaussian noise generator ($D = 0.1, E = 0$) and the BCI signal. $n = 16, \Delta t = 1, t_1 = 0, N = 256.$}
 %  \label{fig_ber_3}
%\end{figure}
\subsection{Evaluating the Number of Sources (Oscillators) of BCI data}

\begin{enumerate}
\item The BCI signal in the form of a real-valued time series of 256 samples was obtained from the author's EEG-based BCI with a sampling rate of 256 Hz~\cite{ref-Aicumene}. The recording corresponded to a calm wakefulness.
\item
Time series unfoldings were obtained for $N = 256, n = 16$.
The scattering matrix was calculated, the eigenvalues of the scattering matrix $W$ were calculated, and the spectrum of the factor analysis of the eigenvectors was estimated. See Fig.~\ref{fig_ber_4}.

\begin{figure}[h!]
    \centering
            \subfigure[Left frontal channel (F1).]{\includegraphics[width=0.49\linewidth]{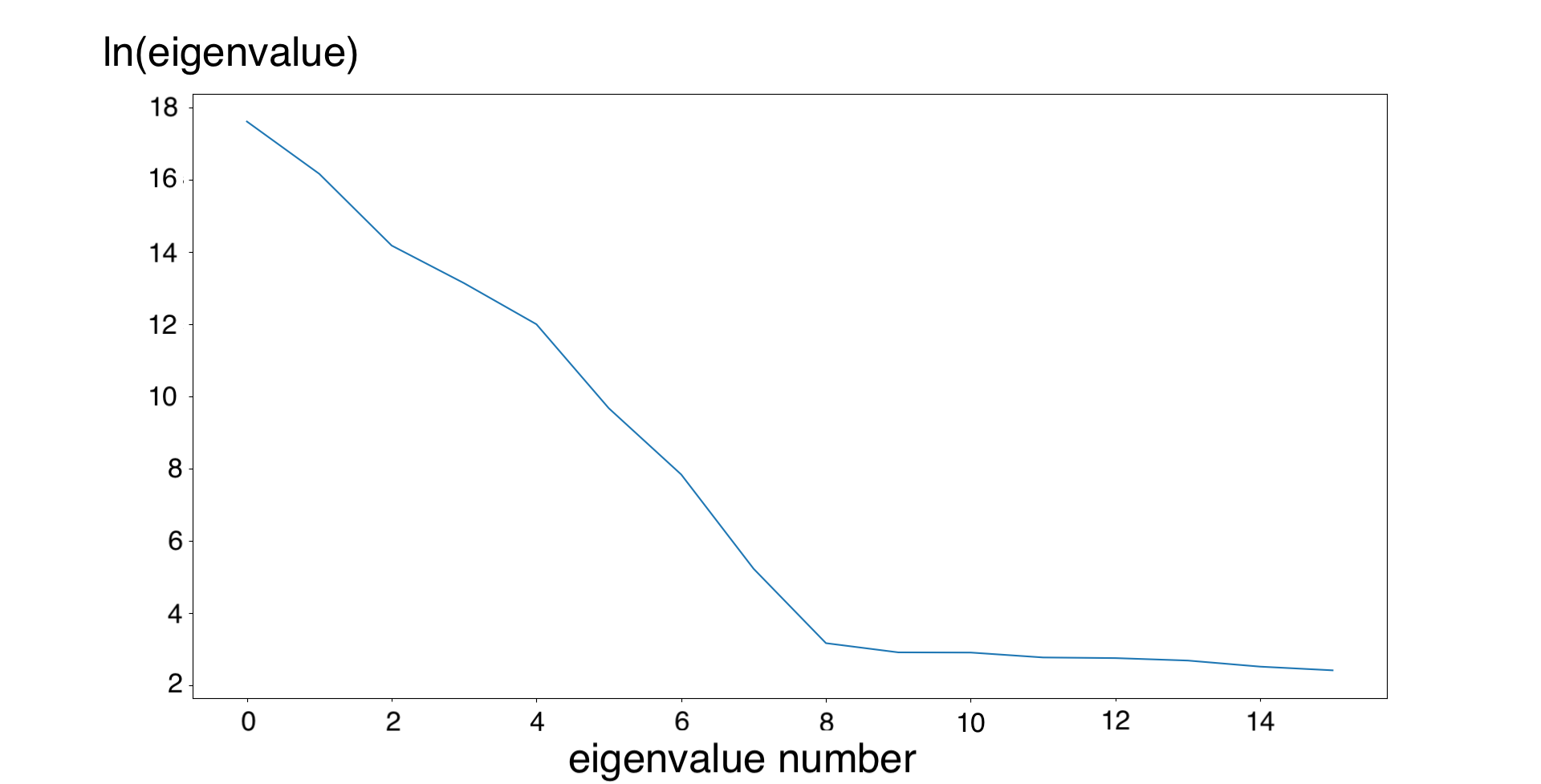}}
    \subfigure[Right frontal channel (F2).]{\includegraphics[width=0.49\linewidth]{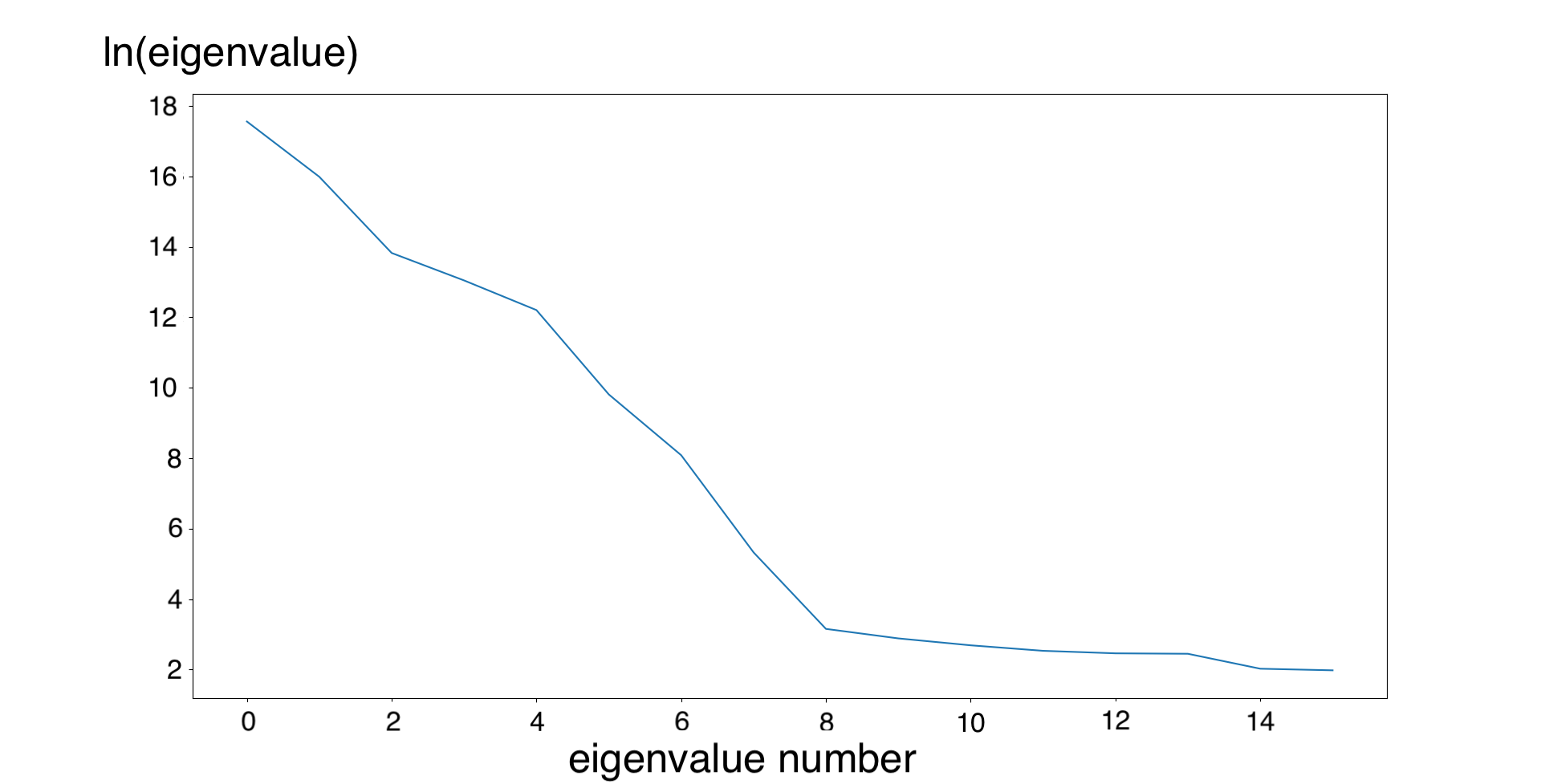}}
    
            \subfigure[Left central channel (C3).]{\includegraphics[width=0.49\linewidth]{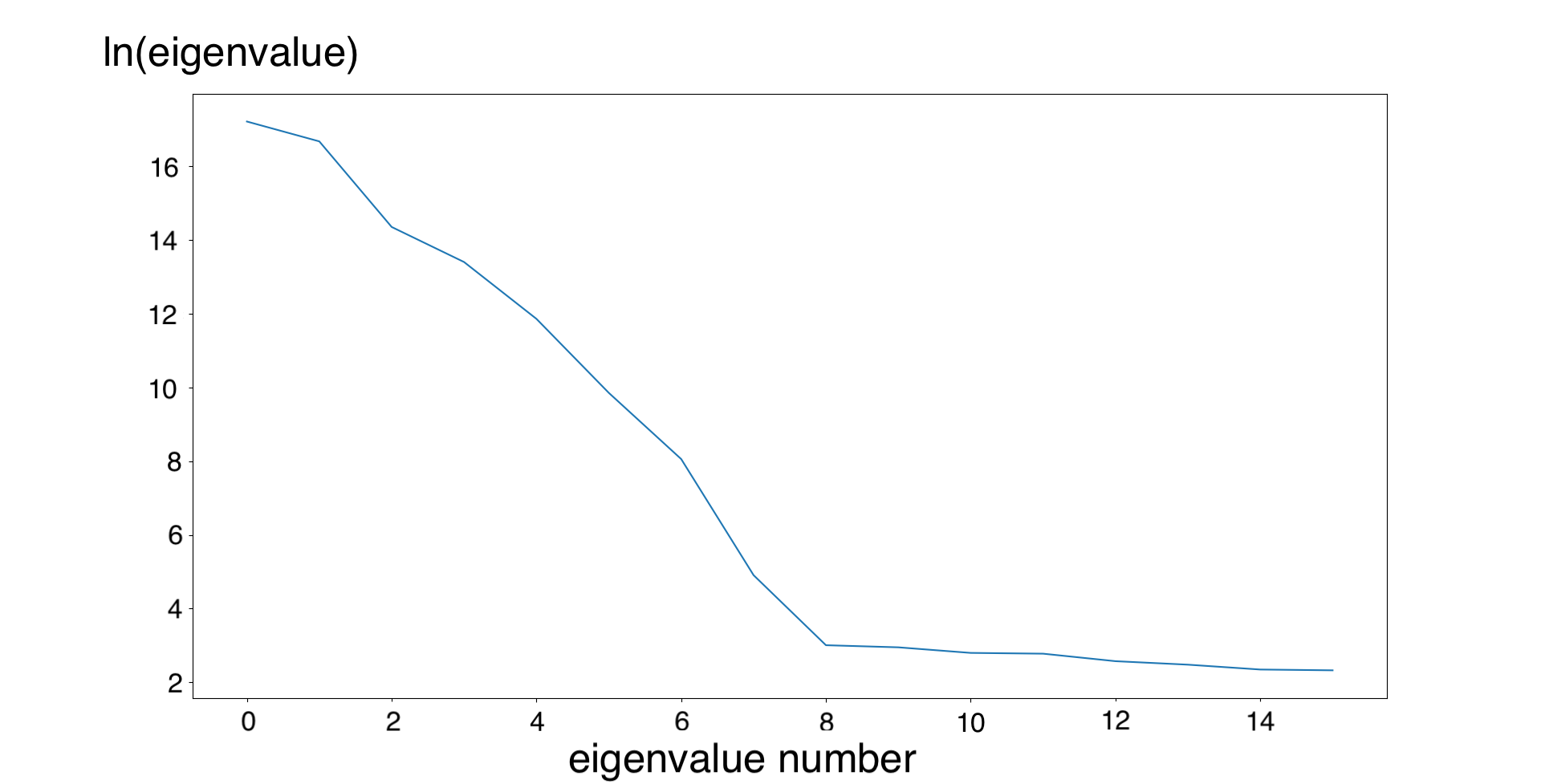}}
    \subfigure[Right central channel (C4).]{\includegraphics[width=0.49\linewidth]{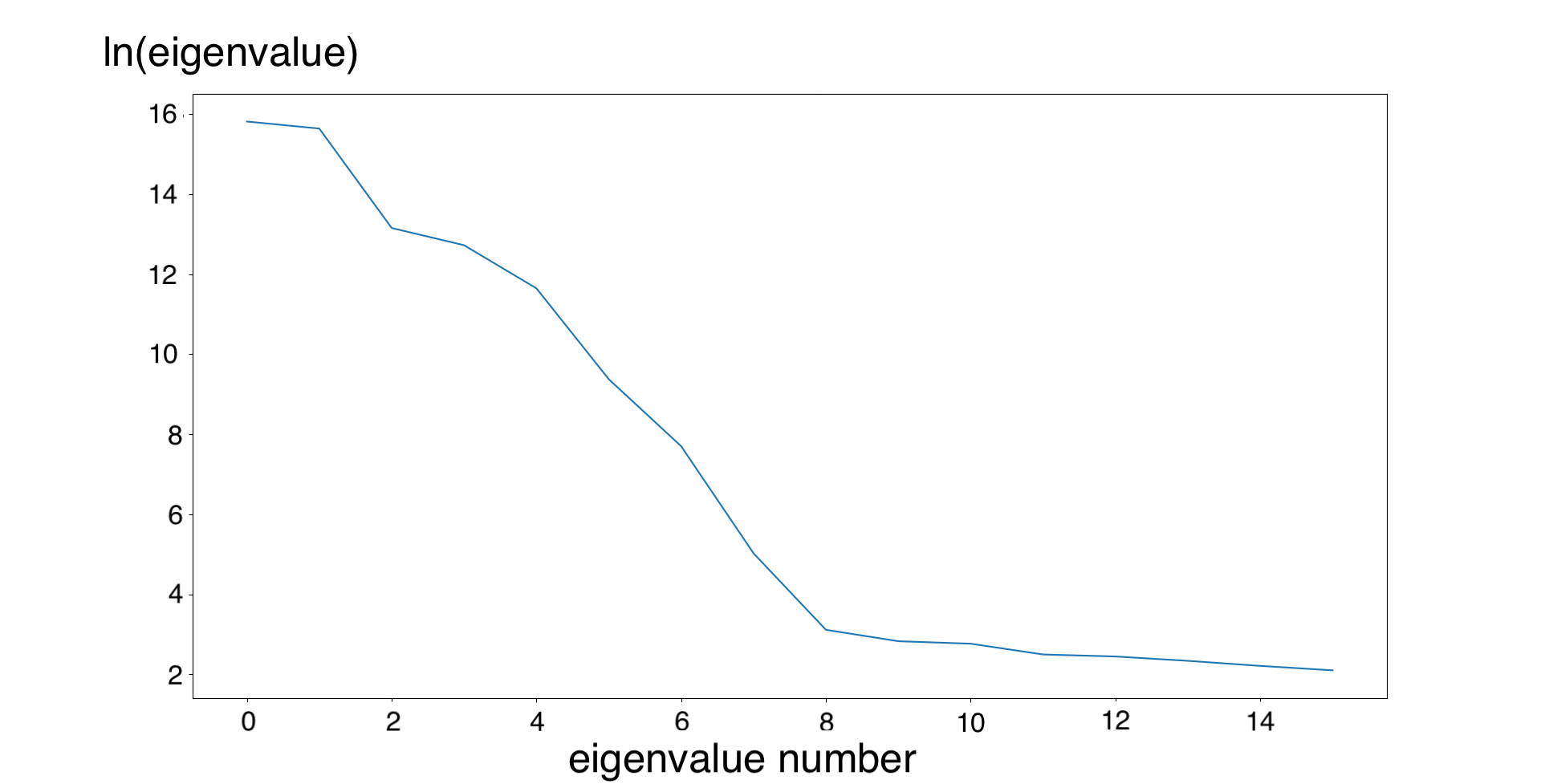}}
    
        \subfigure[Left occipital channel (O1).]{\includegraphics[width=0.49\linewidth]{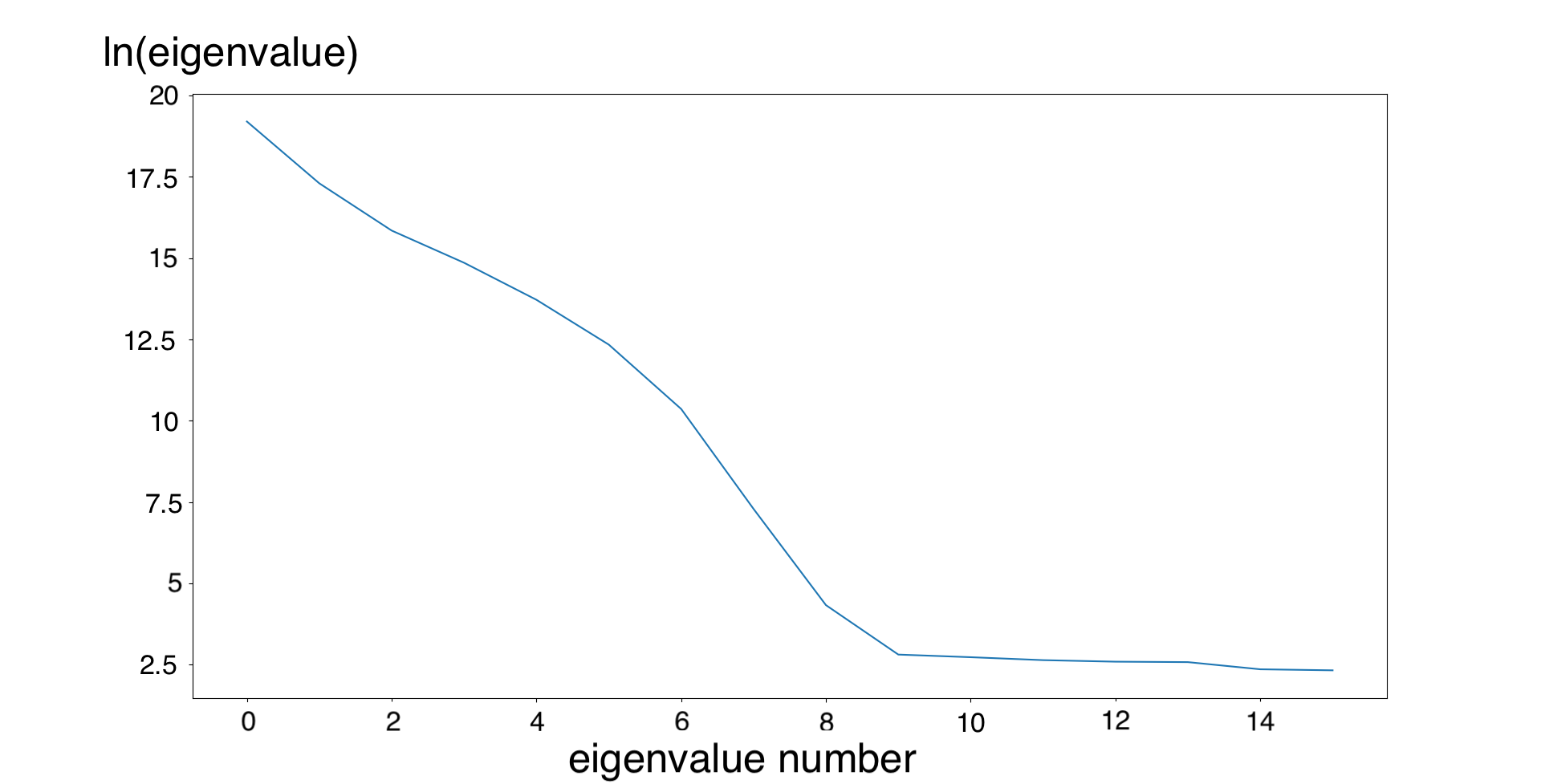}}
    \subfigure[Right occipital channel (O2).]{\includegraphics[width=0.49\linewidth]{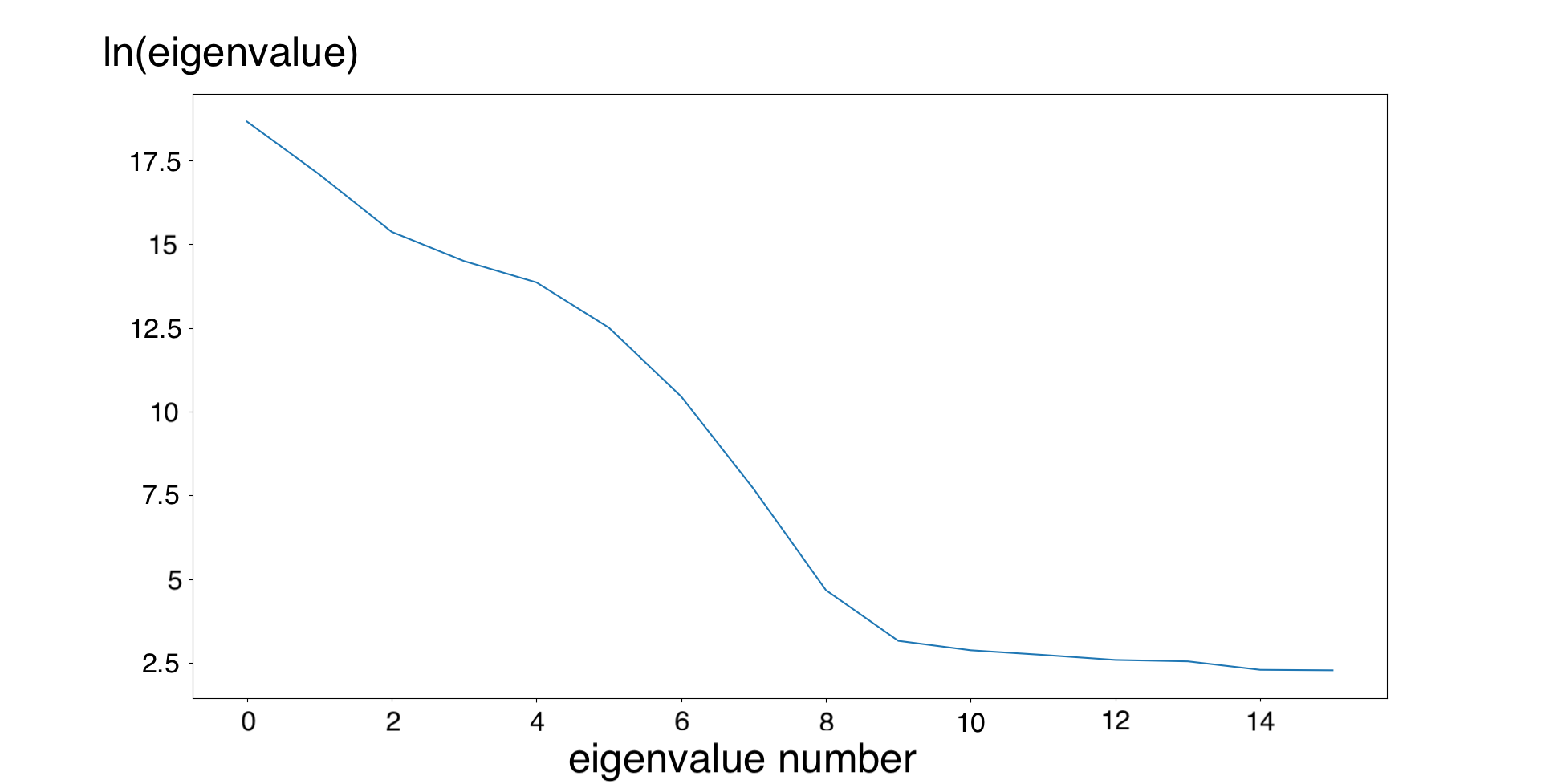}}

    \caption{Logarithmic scale of eigenvalues of the BCI signal unfoldings. $n = 16, \Delta t = 1, N = 256.$ }
    \label{fig_ber_4}
\end{figure}

% \begin{figure}[thpb]
 %    \centering
  %  {\includegraphics[width=0.9\linewidth]{pictures/eigh_F1_.png}}
 %\caption{Logarithmic scale of eigenvalues of BCI data. $n = 16, \Delta t = 1, t_1 = 0, N = 256.$ }
  %   \label{fig_ber_4}
  %\end{figure}

When using the algorithm for evaluating the number of sources, the choice of the threshold for the eigenvalues depends on a priori information about the signal. Such information may be knowledge of the approximate number of oscillators that form the signal, knowledge of the presence of Gaussian noise in the signal, and the results of the previous analysis of similar signals.

Information about the amplitude of the noise in the signal is significant for choosing the threshold value when forming the $\varepsilon$-rank of the unfolding curve.

When working with BCIs, the limiting factor for the accuracy of the signal is the features of the analog-to-digital converter (ADC). In the specification of the BCI ADC, the manufacturer provides information on the noise characteristics. For the data of the author's BCI, the noise is below 2 dB.

From the analysis of the Fig.~\ref{fig_ber_4} we can accept the hypothesis that the information subspace has a dimension of 8. Therefore, the number of oscillators is equal to 4.

%Preliminarily, the first 8 eigenvectors are informative, eigenvectors of smaller amplitude are noise. The number of oscillators is 4.
\item
We filtered the data (See Fig.~\ref{fig_ber_5}) using the subspace $M^r(n, p)$ spanned by 8 eigenvectors for n = 16. In this way, we reduce the dimensionality of the data.
\begin{figure}[h!]
    \centering
            \subfigure[Left frontal channel (F1).]{\includegraphics[width=0.40\linewidth]{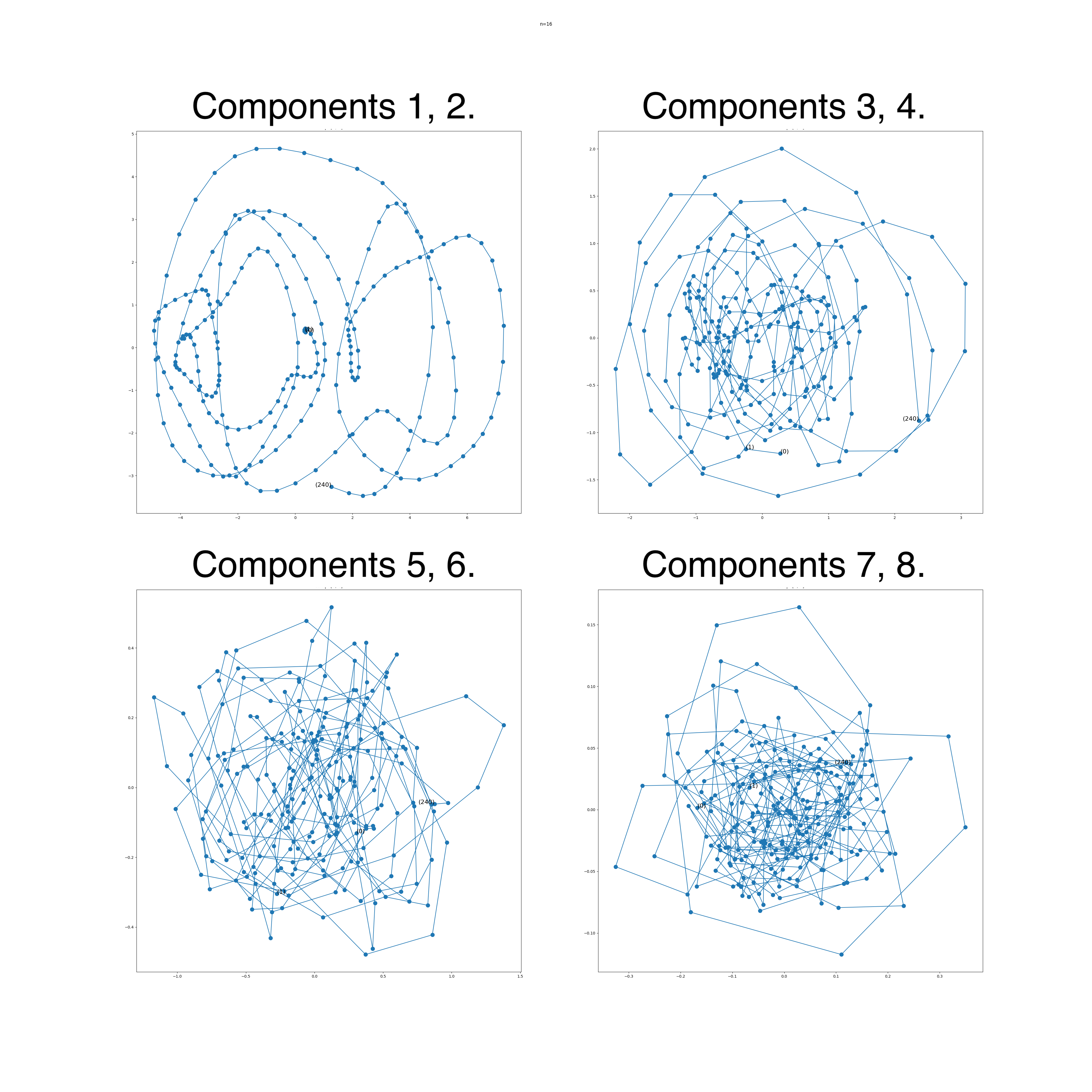}}
    \subfigure[Right frontal channel (F2).]{\includegraphics[width=0.40\linewidth]{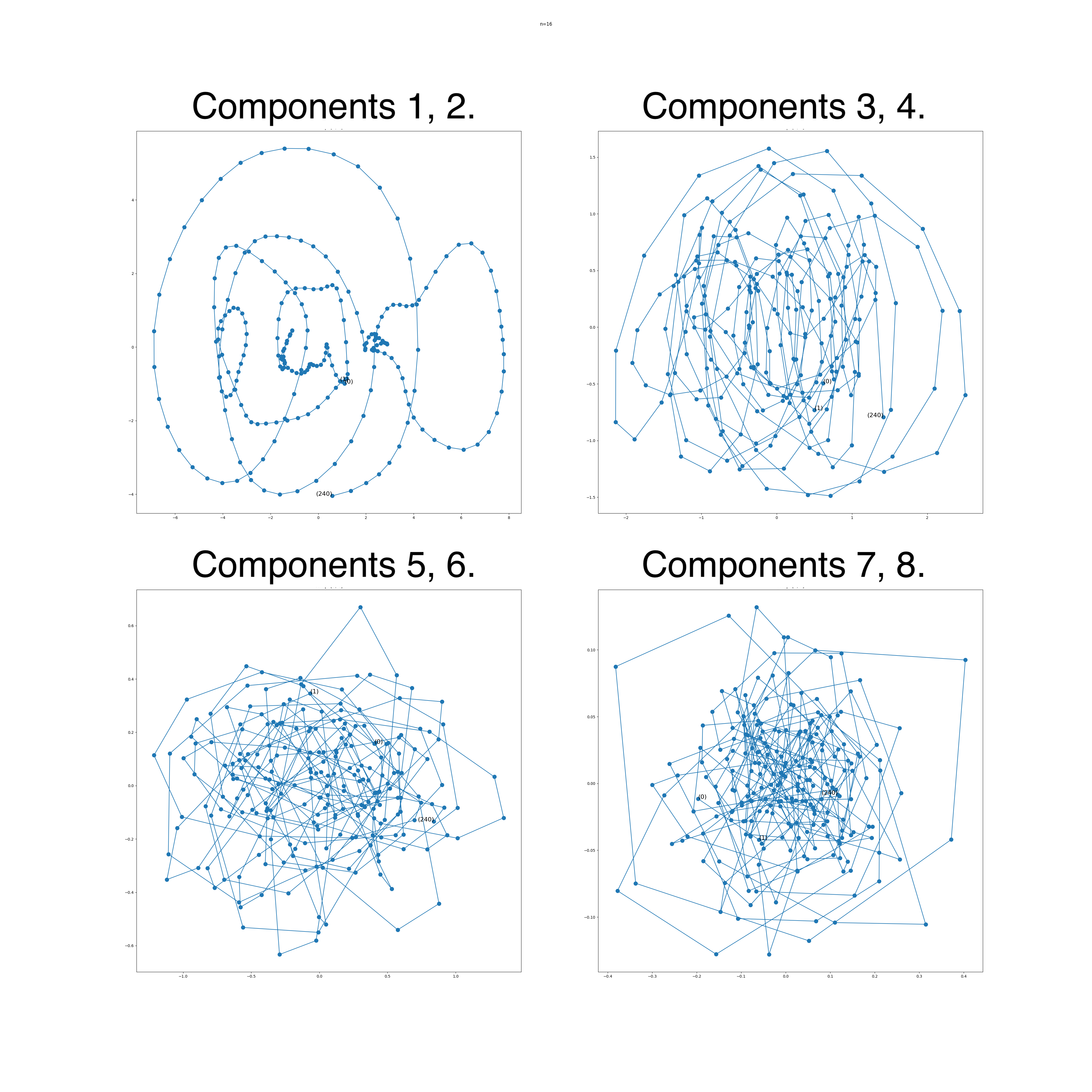}}
    
            \subfigure[Left central channel (C3).]{\includegraphics[width=0.40\linewidth]{N256_co2c0000394_56_C3_pca_n16.png}}
    \subfigure[Right central channel (C4).]{\includegraphics[width=0.40\linewidth]{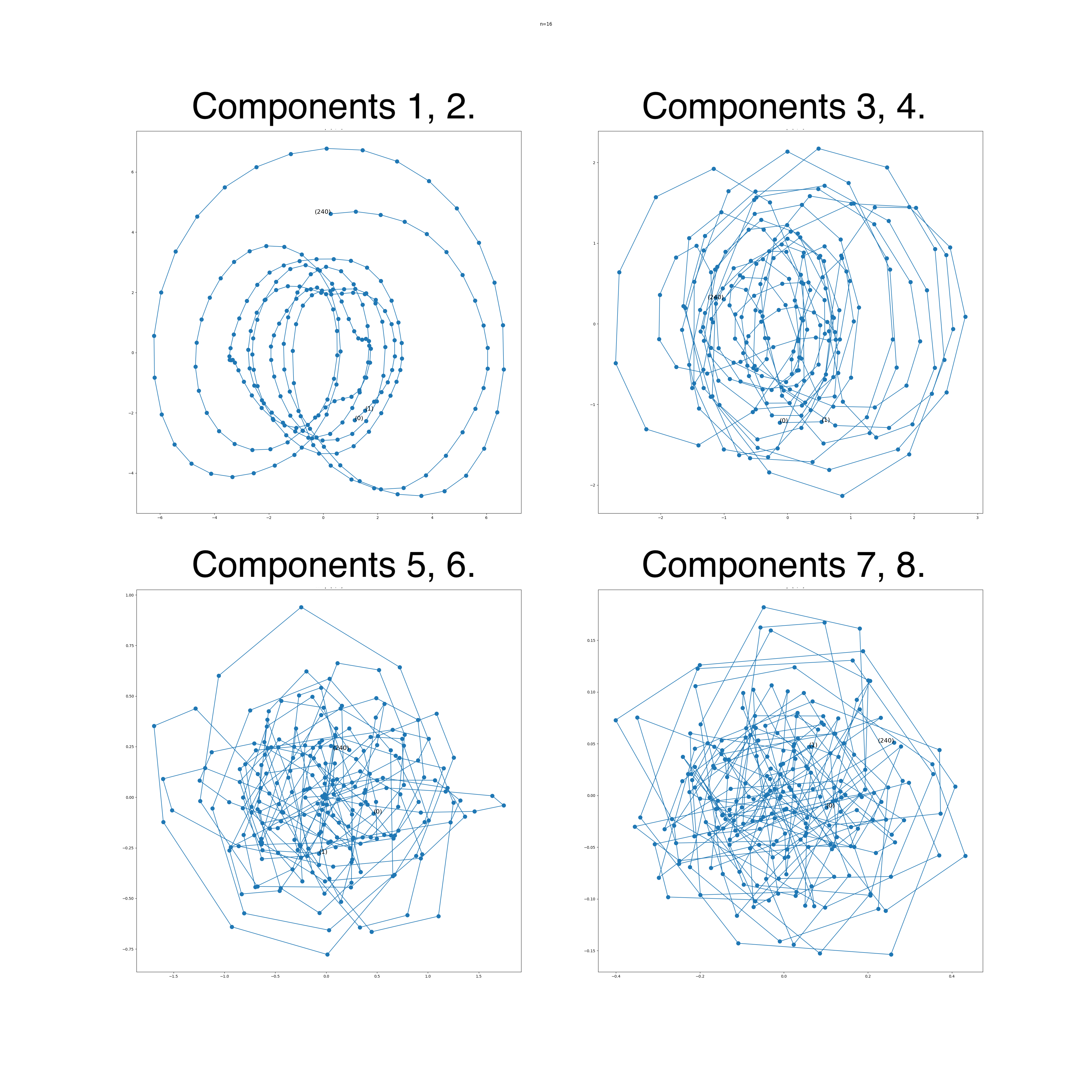}}
        \subfigure[Left occipital channel (O1).]{\includegraphics[width=0.40\linewidth]{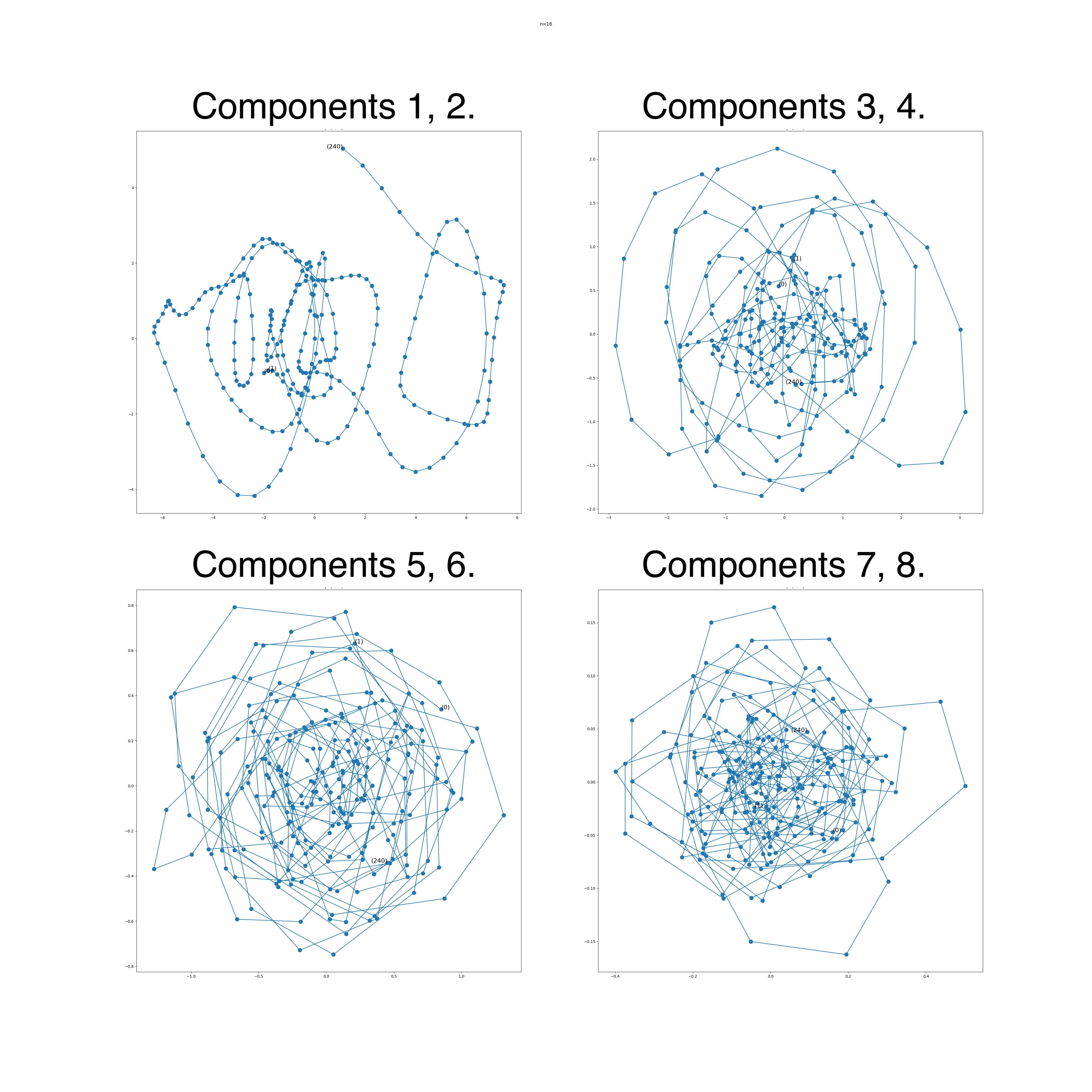}}
    \subfigure[Right occipital channel (O2).]{\includegraphics[width=0.40\linewidth]{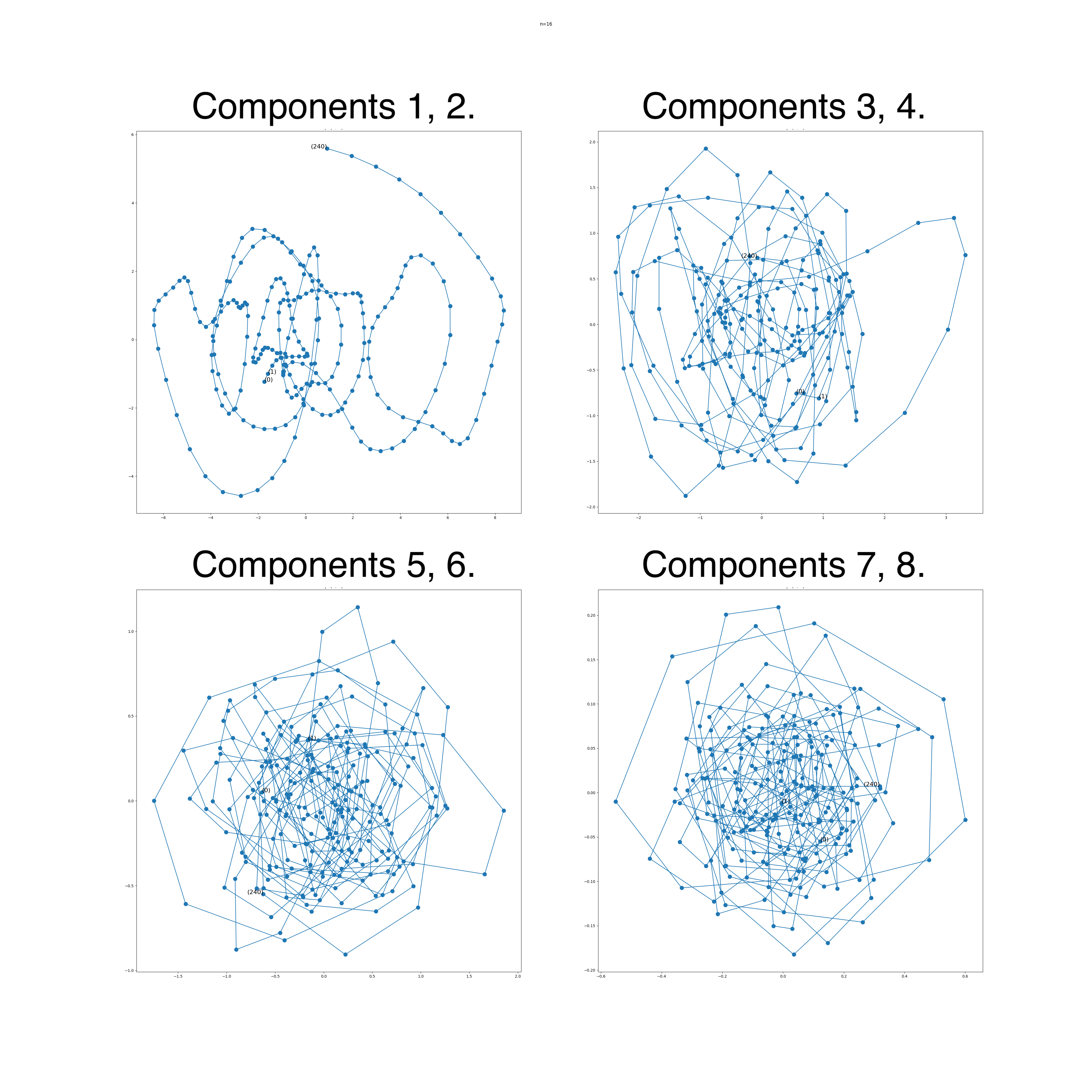}}
    \caption{Projections of the unfoldings onto 1-2, 3-4, 5-6 and 7-8 eigenvectors (components) plans. $n = 16.$}
    \label{fig_ber_5}
\end{figure}

%  \begin{figure}[thpb]
%   \centering
 %          \subfigure[Left frontal channel (F1).]{\includegraphics[width=0.4\linewidth]{pictures/N256_co2c0000394_56_F1_pca_n16.png}}
  % \subfigure[Right frontal channel (F2).]{\includegraphics[width=0.90\linewidth]{pictures/N256_co2c0000394_56_F2_pca_n16.png}}
       %    \subfigure[Left temporal channel (C3).]{\includegraphics[width=0.9\linewidth]{pictures/N256_co2c0000394_56_C3_pca_n16.png}}
%   \subfigure[Right temporal channel (C4).]{\includegraphics[width=0.40\linewidth]{pictures/N256_co2c0000394_56_C4_pca_n16.png}}

%   \caption{Projections of time series unfoldings onto 1-2, 3-4, 5-6 and 7-8 eigenvectors. $n = 16, \Delta t = 1, t_1 = 0, N = 256.$ }
 %  \label{fig_ber_5}
%\end{figure}

We perform the time series reconstruction and evaluate the contribution of each pair of components. See Fig.~\ref{fig_ber_6} and Fig.~\ref{fig_ber_7}.
We compare the original signal with the time series reconstructed by the first pair of components, then by the first four, then by the first six and, at the end, by the first eight components. When reconstructing time series from eight components, the signal residues can be considered noise.

\begin{figure}[h!]

   \centering
%    	\fbox{
       \subfigure[Reconstruction of components 1 and 2.]{\includegraphics[width=0.80\linewidth]{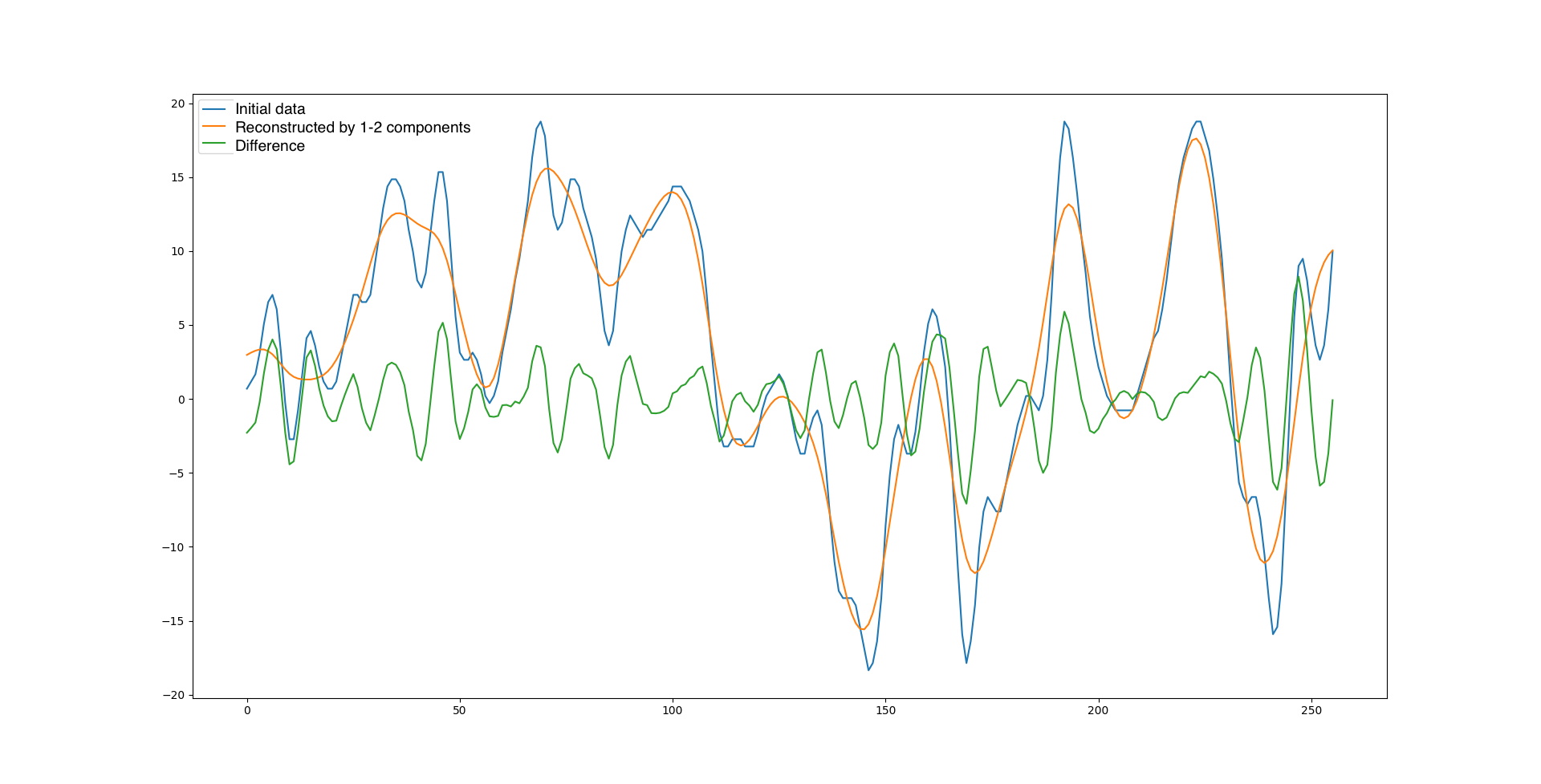}}  
       %}
 
               \subfigure[Reconstruction of components 1 -- 4.]{\includegraphics[width=0.80\linewidth]{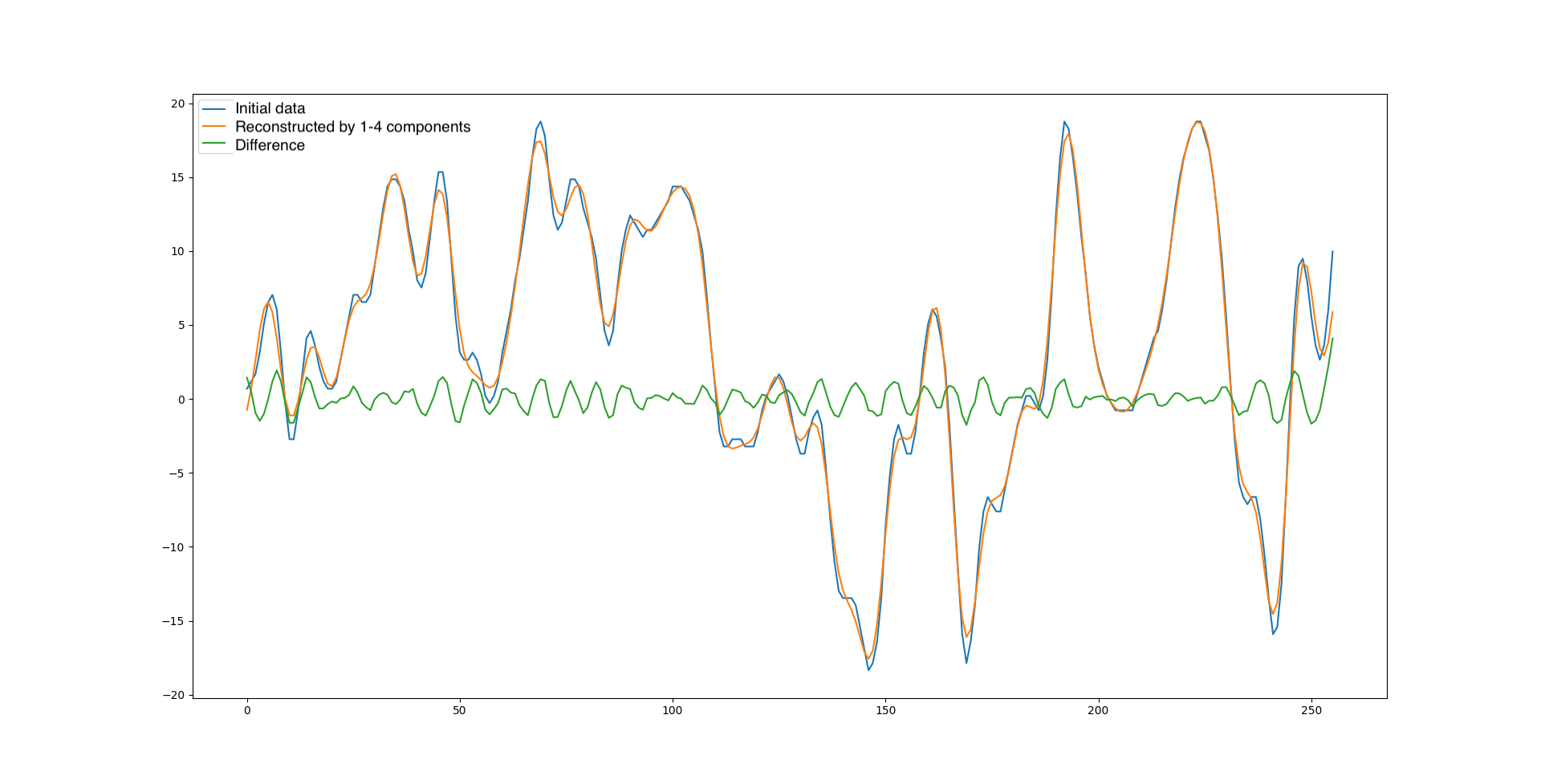}}

                      \subfigure[Reconstruction of components 1 -- 8.]{\includegraphics[width=0.80\linewidth]{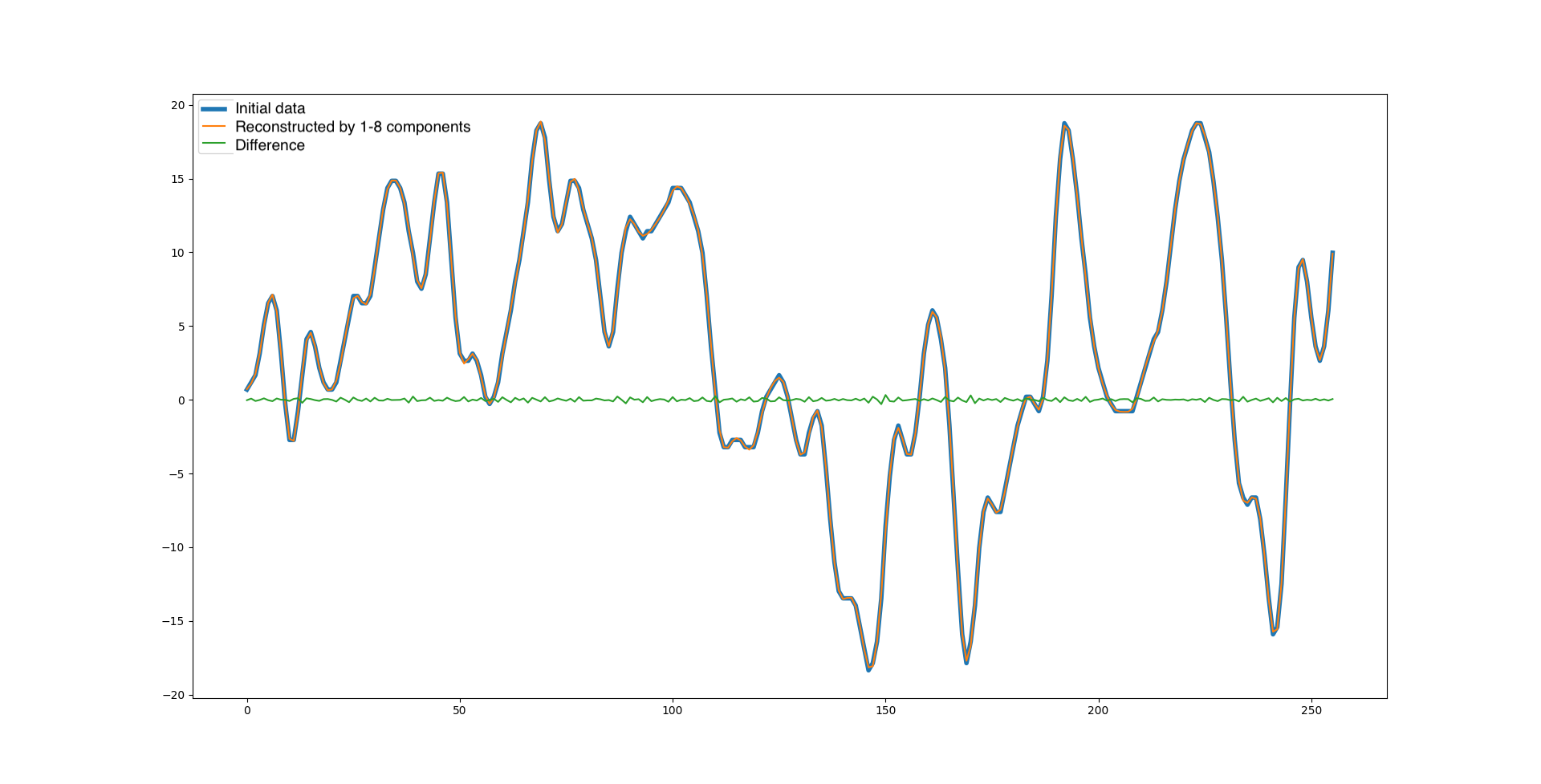}}
                         
   \caption{The original signal of the BCI (channel C3) and the Reconstructed signal component by component. $n = 16, \Delta t = 1, t_1 = 0, N = 256.$}
   \label{fig_ber_6}
\end{figure}

\begin{figure}[h!]
    \centering
            \subfigure[Left frontal channel (F1).]{\includegraphics[width=0.490\linewidth]{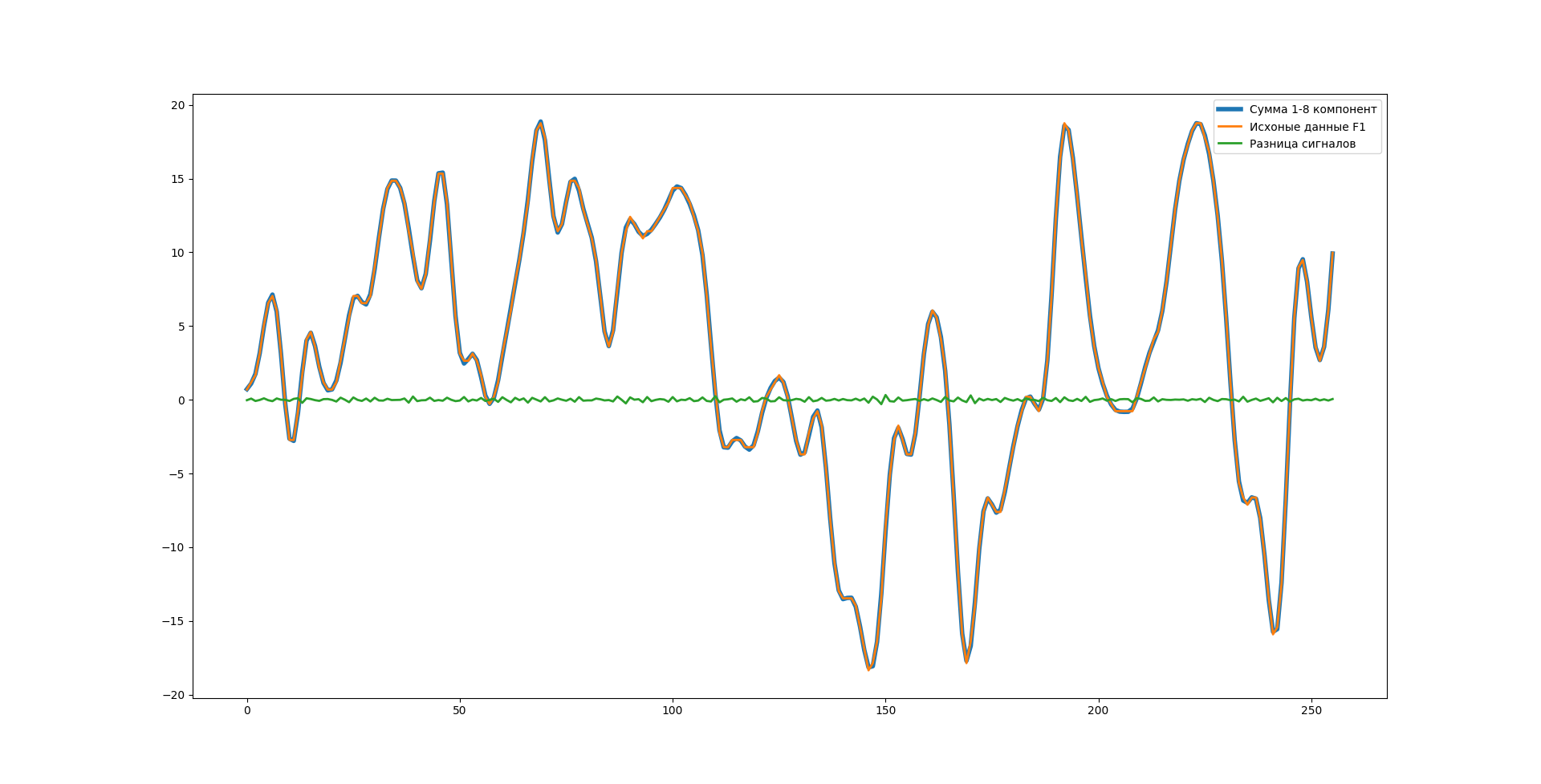}}
    \subfigure[Right frontal channel (F2).]{\includegraphics[width=0.490\linewidth]{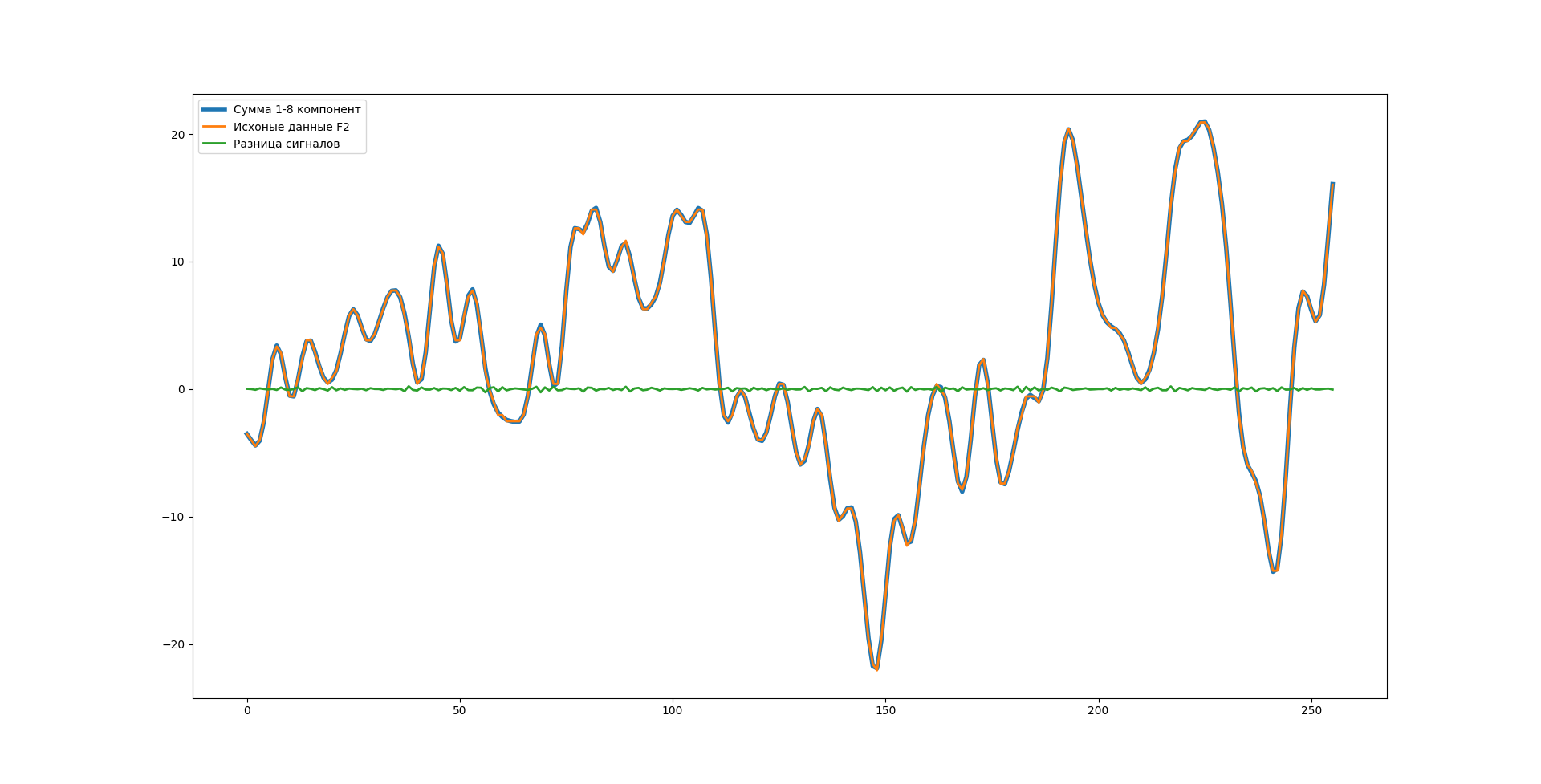}}
    
            \subfigure[Left central channel (C3).]{\includegraphics[width=0.490 \linewidth]{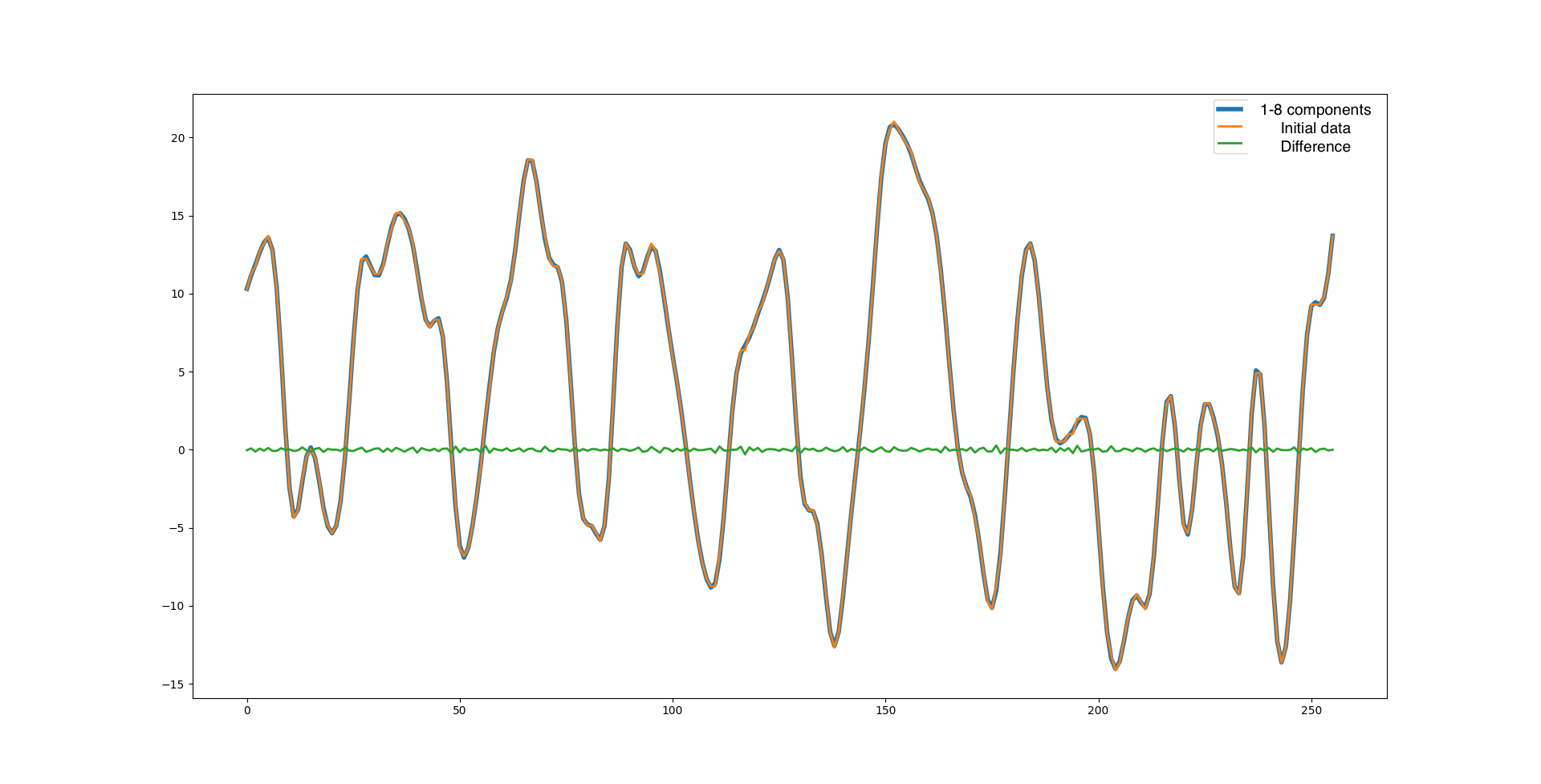}}
    \subfigure[Right central channel (C4).]{\includegraphics[width=0.490\linewidth]{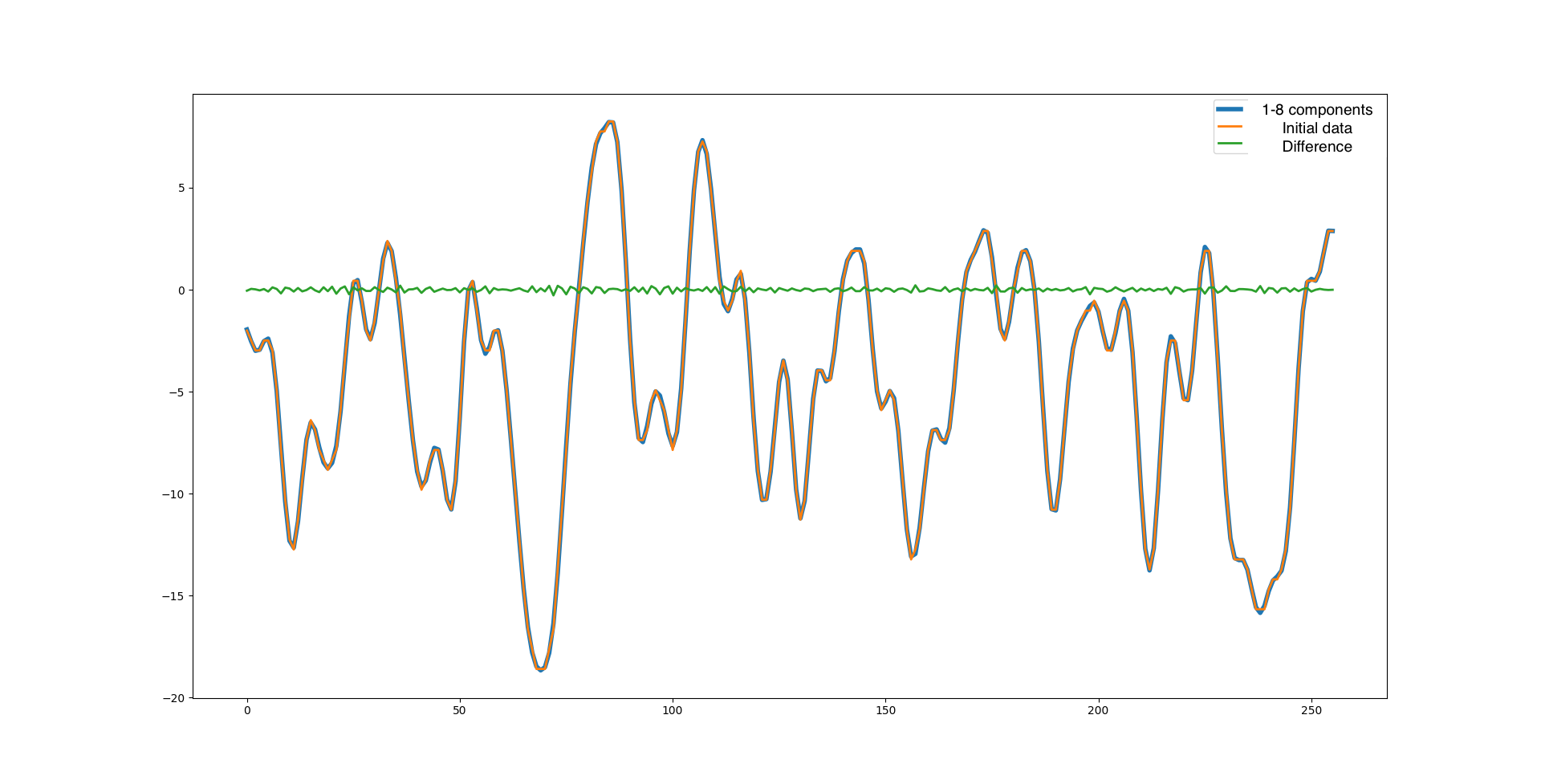}}
    
        \subfigure[Left occipital channel (O1).]{\includegraphics[width=0.490\linewidth]{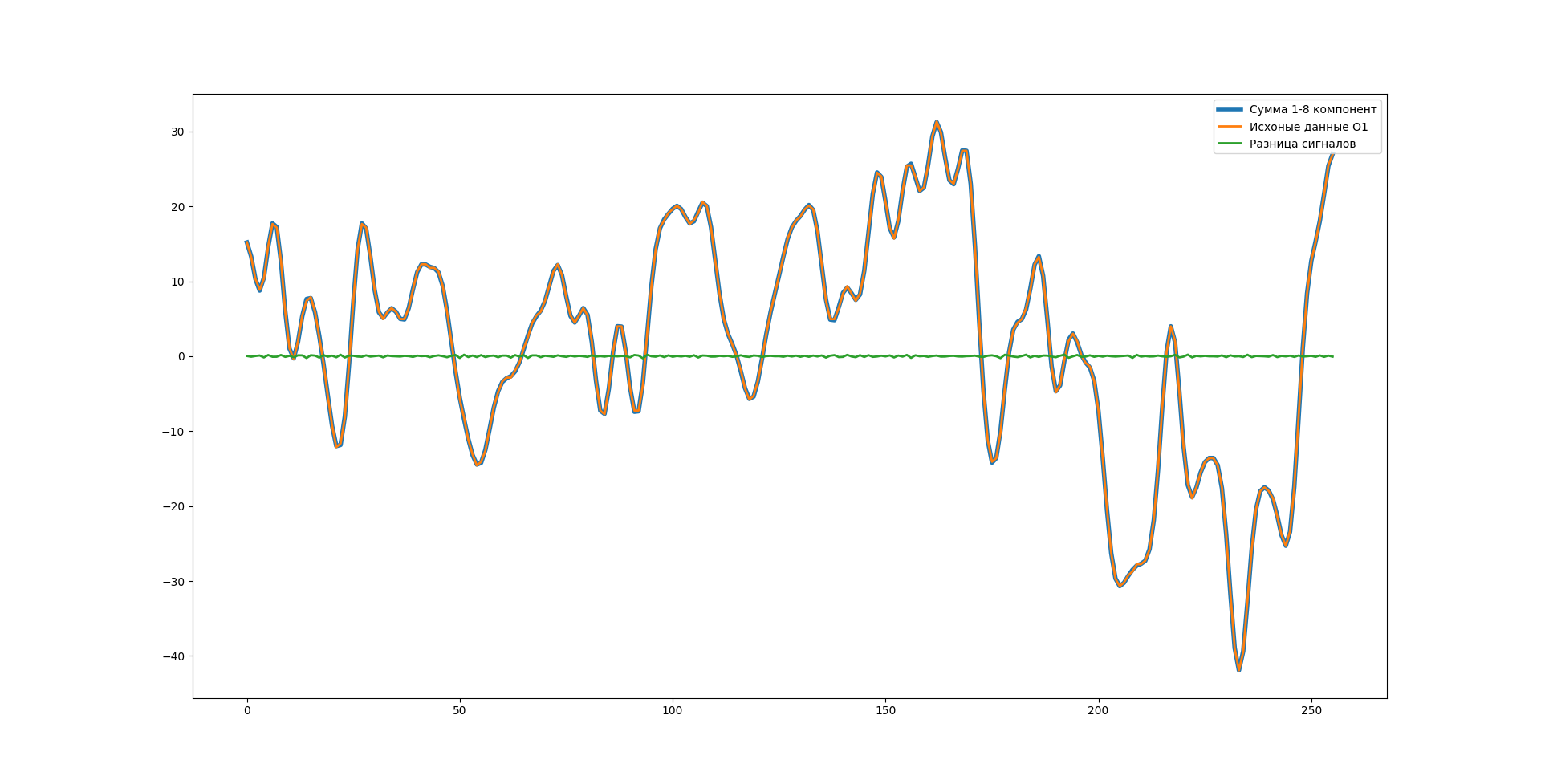}}
    \subfigure[Right occipital channel (O2)]{\includegraphics[width=0.490\linewidth]{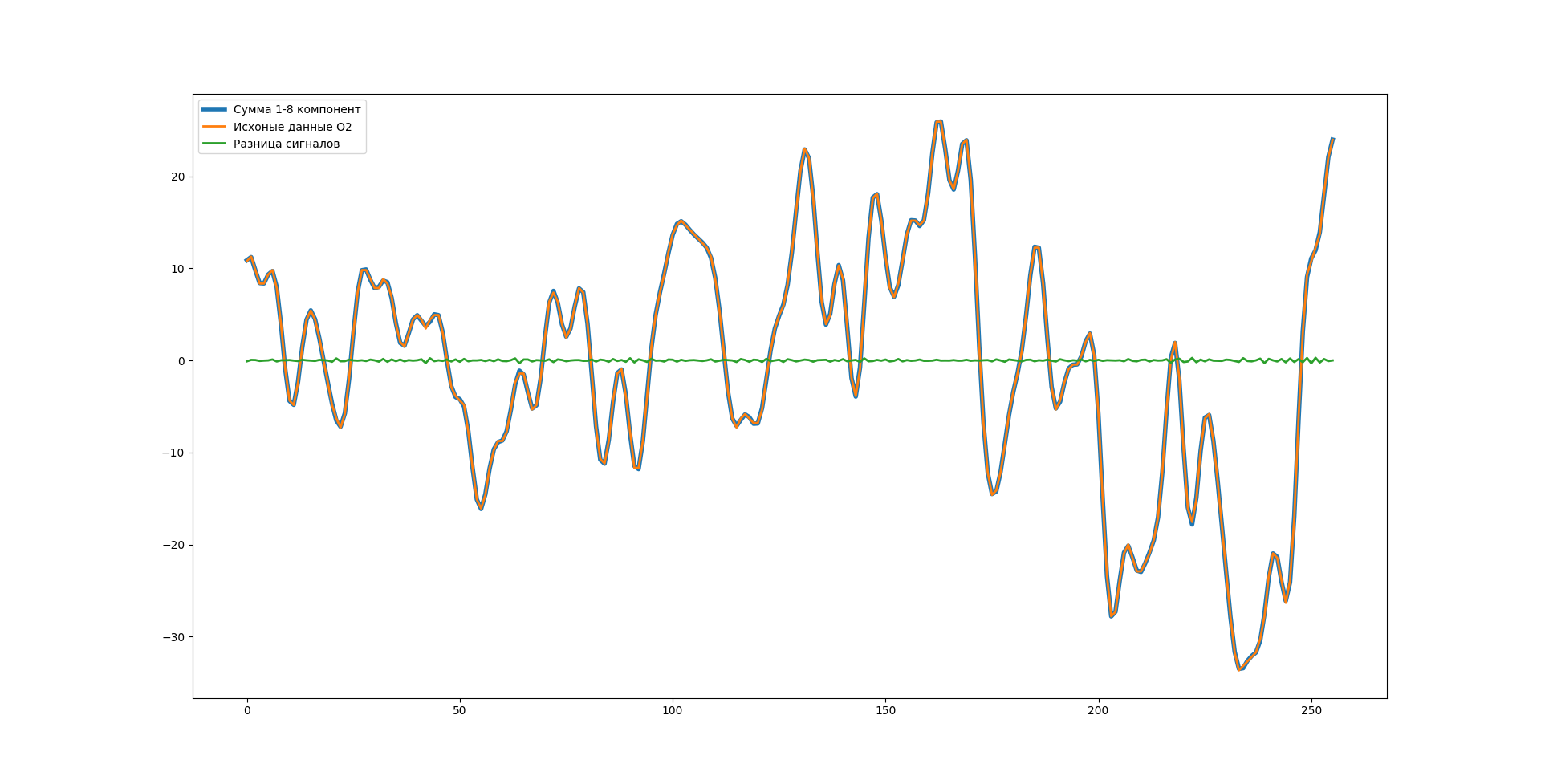}}

    \caption{The original time series of the BCI and the reconstructed by 8 components. Channels: F1, F2, C3, C4, O1, O2 (correspond to the international 10-20 scheme).  The original time series is colored blue, the reconstructed time series from eight components is colored orange, and the signal residues (noise) is colored green. $n = 16, \Delta t = 1, t_1 = 0, N = 256.$ }
    \label{fig_ber_7}
\end{figure}

For BCI data, the time series unfolding is located with high accuracy in the eight-dimensional space. Therefore, the required number of oscillators of this signal, according to our model, corresponds to 4.
\end{enumerate}

\section{CONCLUSIONS}
In this work we use the polyharmonic model of brain signals detected by the sensors of a non-invasive BCI.

The work provides the necessary introduction to the theory of the time series unfolding method.

The paper proposes an algorithm for solving the problem of estimating the number of active brain oscillators based on the results of the analysis of the unfolding of model polyharmonic functions.

The results are demonstrated using data from the author's non-invasive BCI.

The next publication will describe methods for analyzing a set of all signals detected by the BCI; algorithms for estimating the number of active clusters that form the set of signals detected by the BCI; algorithms for reconstructing the mutual arrangement of active signal sources. The results of these algorithms will be demonstrated using the data of the author's BCI.

%\subsection{Citations}
%Citations use \verb+natbib+. The documentation may be found at
%\begin{center}
%	\url{http://mirrors.ctan.org/macros/latex/contrib/natbib/natnotes.pdf}
%\end{center}

%Here is an example usage of the two main commands (\verb+citet+ and \verb+citep+): Some people thought a thing \citep{kour2014real, keshet2016prediction} but other people thought something else \citep{kour2014fast}. Many people have speculated that if we knew exactly why \citet{kour2014fast} thought this\dots

%\section{REFERENCES}
%\bibliographystyle{unsrtnat}
%\bibliography{references}  %%% Uncomment this line and comment out the ``thebibliography'' section below to use the external .bib file (using bibtex) .

%\renewcommand{\bibname}{REF}
%\bibliography{publications}
%\bibliographystyle{plain}
\renewcommand{\refname}{REFERENCES}
%\newpage

\end{document}